\documentclass[8pt]{article}
\usepackage{newinutile}
\usepackage{everyshi}
\usepackage[utf8]{inputenc}
\usepackage{amsmath,amssymb,amsthm,amsfonts}
\usepackage{epsfig,graphics,color,calc,graphicx}
\usepackage{subfigure}
\graphicspath{{Figures/}}
\usepackage{comment}
\usepackage{pgf,tikz}
\usepackage{mathrsfs}
\usepackage{color}
\usetikzlibrary{arrows}
\usepackage{cancel}
\usepackage{mathrsfs}
\usepackage{hyperref}
\hypersetup{colorlinks=true,linkcolor=blue}
\usepackage{placeins}
\newcommand\be{\begin{equation}}
\newcommand\ee{\end{equation}}
\newcommand\bea{\begin{eqnarray}}
\newcommand\eea{\end{eqnarray}}

\newcommand\zw{\omega}
\newcommand\zl{\lambda}
\newcommand\zb{\beta}

\newcommand\zt{\tau}
\newcommand\ensavg{\left\langle e^{-\beta W_J} \right\rangle}
\DeclareMathOperator\erf{erf}

\def\lsim{\,\lower2truept\hbox{${< \atop\hbox{\raise4truept\hbox{$\sim$}}}$}\,}
\def\gsim{\,\lower2truept\hbox{${> \atop\hbox{\raise4truept\hbox{$\sim$}}}$}\,}

\newlength{\pecettawidth}
\newcommand{\pecetta}[1]{
	\medskip
	\begin{equation}gin{center}
		\framebox{
			\begin{array}gin{minipage}{\pecettawidth}\texttt{Commento.}\ #1\end{minipage}
		}
	\end{center}
	\medskip
}

\begin{document}
	%%% Font: commenta la riga che segue se vuoi i font standard
	%\fontfamily{ppl}\selectfont
	%\logo
	\title{{Finite reservoirs and irreversibility corrections to Hamiltonian systems statistics
 %On the protocol and finite size effects dependence  of the Jarzynski Equality
 }}
		
\author{Matteo Colangeli}
	\email{matteo.colangeli1@univaq.it}
	\affiliation{Dipartimento di Ingegneria e Scienze dell'Informazione e
		Matematica,
		Universit\`a degli Studi dell'Aquila, via Vetoio, 67100 L'Aquila, Italy.}
	
	\author{Antonio Di Francesco}
	\email{antonio.difrancesco5@graduate.univaq.it}
	\affiliation{Dipartimento di Ingegneria e Scienze dell'Informazione e
		Matematica,
		Universit\`a degli Studi dell'Aquila, via Vetoio, 67100 L'Aquila, Italy.}

	\author{Lamberto Rondoni}
	\email{lamberto.rondoni@polito.it}
	\affiliation{Dipartimento di Scienze Matematiche,
		Politecnico di Torino, Corso Duca degli Abruzzi 24, 10129 Torino, Italy \\
  INFN, Sezione di Torino, Via Pietro Giuria 1, 10125 Torino, Italy \\
  ORCID: 0000-0002-4223-6279 }
	%\thanks{}
	
\abstract{We consider several Hamiltonian systems perturbed by external agents, that preserve their Hamiltonian structure. We investigate the corrections to the canonical statistics resulting from coupling such systems with possibly large but finite 
reservoirs, and from the onset of processes breaking the time reversal symmetry. We analyze exactly solvable oscillators systems, and perform simulations of relatively more complex ones. This indicates that the standard statistical mechanical formalism needs to be adjusted, in the ever more investigated nano-scale science and technology.
In particular, the hypothesis that heat reservoirs be considered infinite and be described by the classical ensembles is found to be critical when exponential quantities are considered, since the large size limit may not coincide with the infinite size canonical result. Furthermore, 
process-dependent emergent irreversibility affects ensemble averages, effectively frustrating, on a statistical level, the time reversal invariance of Hamiltonian dynamics, that is used to obtain numerous results.}

	%\pacs{05.20.Dd}
	
	%\preprint{Appunti: \today}
	
	%\vfill
	%\noindent
	%\textbf{MSC2000:} 82B28; 82B44; 60K45.
	%PACs: 02.50.Ga, 05.40.Fb, 05.10.Ln
	\maketitle
	
	\section{Introduction}
	\label{s:intro}
	\par\noindent
%The statistical mechanics approach to  thermodynamic phenomena is based on the class of Hamiltonian particle systems.
The validity of the canonical ensemble is universally accepted to compute macroscopic quantities of systems in equilibrium at a given temperature $T$, as averages of phase space functions. The corresponding  formalism assumes that heat reservoirs are infinitely large, and that measurement times are exceedingly longer than the characteristic times  of the microscopic events. 
Such mathematical idealizations yield a highly successful theory describing a vast range of macroscopic phenomena. The separation between microscopic and macroscopic scales is indeed  sufficiently wide for calculations of quantities of thermodynamic interest.
Nevertheless, there are various reasons for investigating the applicability of the canonical framework to non-standard observables. For instance, exponentials of microscopically expressed variables appear in Bennett's formulae for the free energy \cite{bennett1976efficient}, in
Widom's relation \cite{widom1963some}, 
Zwanzig's relation \cite{zwanzig1955high}, and in the more recent Jarzynski \cite{jarzynski1997nonequilibrium} and Crooks relations \cite{crooks1999entropy}.
Furthermore, current science and technology deal with small  systems and fast processes, as well as with quantities not immediately interpretable in thermodynamic terms, as in the case of anomalous enegy transport
\cite{giberti2019temperature,giberti2019n,dematteis2020coexistence}. 
Therefore, finite size effects and lack of ergodicity may turn important. Indeed, standard thermodynamic properties of macroscopic objects only require a proper characterization of the bulk of the relevant probability distributions, not of their tails. 
On the other hand, an accurate characterization of the tails of the relevant probability distributions becomes necessary when dealing with observables that
get a substantial contribution from such tails. Then, the fact that thermal baths are necessarily finite and that experiments may last very short times may require particular attention. 

%Thermodynamics holds irrespective of its microscopic interpretation, because it deals with large systems and simple observables, measured over macroscopic intervals of time. However, 
%Consequently, statistical mechanics %provides more than an interpretation of otherwise known results. Not only it 
%has first led to profound results and predictions that could have not been drawn from thermodynamics alone.

%The results derived from the usual equilibrium ensembles may then require revision. 

In this work we take the quantity used in the Jarzynski Equality (JE) as a paradigmatic example of topical non-standard observables. It is worth recalling that the time reversal symmetry of the microscopic dynamics \cite{Lax,Sachs,RB,RQ} is essential for the derivation of the JE, which belongs to a class of results, known as \textit{Fluctuation Relations}, strongly relying on the time reversibility of the microscopic dynamics, see e.g. \cite{Gall,ECM,Gasp}. More generally, the time reversal symmetry turns out being a standard ingredient of a large variety of statistical mechanical results, including the Onsager Reciprocal Relations \cite{Ons1,Ons2}, Fluctuation-Dissipation Theorem  and the Green-Kubo relations \cite{Callen2,Kubo1,Kubo2}, and applications to magnetic systems \cite{Casimir}.
In works such as Ref.\cite{ciccotti2022jarzynski}
it was found that certain nanoscopic Hamiltonian systems violate the JE, although formally amenable to analysis within the canonical framework, that yields the JE as an exact relation 
\cite{jarzynski1997nonequilibrium}. 
%In Ref.\cite{ciccotti2022jarzynski}, a nanoscopic crystal undergoes plastic deformations, when an indenter pentrates sufficiently deeply in it. 
In the case of Ref.\cite{ciccotti2022jarzynski},
the failure was caused by the emergence of irreversibility 
%of Hamiltonian systems at the nanoscale, 
due to a process dependent nonequilibrium effect, and not to the large number of degrees of freedom. %invoked in the usual picture of irreversibility.
%, is an interesting discovery. 
At the same time, highly nonequilibrium processes do not prevent the validity of the JE in {\em e.g.}\ 1-dimensional systems described by an overdamped Langevin equation \cite{SeifertReview}.
That this may be the case is clear in the words of, {\em e.g}\ Fermi \cite{Fermi} or Callen \cite{Callen},
who state, in practice, that the ensembles work if the observation times suffice for the observables of interest to have thouroughly explored their range. Khinchin then adds that this is easy to obtain, for the observables of interest typically have a small range \cite{Khinchin}.
In all instances, the state of the system is required to be stationary, or very slowly evolving with respect to the observation times.

The above considerations are topical, given the rapid development of bio- and nano-technologies, which deal with small systems. Apart from being small,  such systems are often briskly driven by external agents, so that thermal baths (even if effectively infinite) only express limited energy, and the deterministic thermodynamic laws must often be replaced by statistical laws. Certainly, some experiments of bio- and nano-technological interest intentionally take very short times, so that only a small part of a thermal bath is effectively involved.
%As noted above, the observables of interest require an accurate characterization of the tails of their probability distributions. 
%As an example of such cases, one can think of the averaging of exponential functions of the phase variables when measured with Gaussian distributions, where clearly one of the tails of the distribution is weighted sensibly more than the other. 
This poses the question, when computing ensemble averages, about proper approaches to the  finiteness of the bath or, in other terms, to the restriction to finite subsets of phase space. %when dealing with canonical measures. 
%In situations like these, ensembles results must be properly scrutinized. 
%Indeed, no thermal bath can really be infinite.  %As shown in \cite{ciccotti2023} through the analysis of an explicitly solvable random walk model, the correction due to finite bounds in the integration of a Gaussian probability density does not vanish in the infinite integration interval limit when the observable is of exponential type.
 
%This is the case, for instance, in the experiments of interest in the JE literature, that can typically sample only a limited region of the available state or phase space. 

In %view of these considerations, in 
this paper we thus
analyze the finite size effects on the statistics concerning simple Hamiltonian systems, subjected to various external drivings. We start by briefly reviewing the derivation of the JE in Sec.\ \ref{s:TheJE}. 
In Sec.\ \ref{s:models_methods}, three simple mechanical models are introduced to illustrate the 
%main ideas and methods considered in this paper, and 
onset of finite size effects that lead to violations of the JE, highlighting some of the limitations of the canonical statistic. In particular it is shown that the speed of of the protocol or the frequency of periodic drivings resonating with the system proper frequency may wildly enhance the protocol dependence, violating of up to 110\% the JE. We also highlight the fact that analagous results are obtained for infinite baths at small temperatures.

In Sec.\ \ref{s:idealgas_box_expansion}, we consider a model mimicking the adiabatic expansion of an ideal gas, and also describe the validity of the JE in the presence of a protocol-dependent device concluding that the occurrence of an irreversible phenomenon (such as the free expansion of a gas) can invalidate the statistical description of a particle system through the canonical formalism. 

Conclusions are drawn in Sec.\ \ref{s:discussion}, where we also anticipate future developments. The Appendices give the details of some analytical calculations reported in the main text.
%of interest varies in time till it returns to its initial form.  
%According to the canonical calculations that yield the JE, the same averages should be obtained, whatever process is performed, as long as initial and final Hamiltonians are the same.
%One application of the classical ensembles that results most useful and clear, from this point of view, is afforded by the \textit{Jarzynski Equality} (JE). Therefore, we mainly focus on it, although the message is of general interest. 
%We address the following questions: 1) what is the phenomenology at finite ensemble size?
%2) does increasing the energy that a bath can exchange make the result  closer to the theoretical prediction for infinite baths?

%For the statistical analysis to be reliable, it suffices that averages be computed over a volume of initial phases larger than the sampled one, the statistics should be safe, while only a finite domain is considered. 

\section{Derivation of the Jarzynski equality}
	\label{s:TheJE}
    \par\noindent
A well-known example involving both the canonical ensemble and exponential variables is the Jarzynski Equality (JE), which offers a useful playground to highlight the role of finite size effects on the statistics of thermodynamics quantities in the canonical framework. The JE has been derived for both stochastic, and deterministic systems. We focus on the second, which concerns a system S made of $N$ particles, initially in equilibrium with a bath B at temperature $T$. The system may interact with an environment, E, also initially in equilibrium with B. The Hamiltonian of system and environment, denoted by S+E, is assumed to take the following form:
\be
    {\cal H}(\mathbf{x},\mathbf{v};\zl) =H_S(x_S,v_S; \lambda) + H_E(x_E,v_E) + h_I(\mathbf{x},\mathbf{v})
\label{JEham} 
    \ee
    where $\zl$ is a parameter controlled by an external agent, $(\mathbf{x},\mathbf{v})=(x_S,v_S,x_E,v_E)$ are the position and velocities of S and E, as indicated by the subscripts, $H_S$, $H_E$ and $h_I$ are, respectively, the energy of S, the energy of E and the energy of their interaction.
The initial distributions of coordinates and momenta of S+E, which is in equilibrium
at temnperature $T$, is given by the canonical ensemble:
\be
    P_0(\Gamma)={1 \over Z_0} e^{-\beta {\cal H}(\Gamma; A)} \, , \quad \beta = \cfrac{1}{k_{_B} T}
    \label{CanEns}
    \ee
where $k_B$ is the Boltzmann's constant, $\Gamma=(\mathbf{x},\mathbf{v})$ is one  configuration of S+E, and $Z_0$ is the initial canonical partition function. 

At time $t=0$, this system is  isolated from the bath, and driven by an external agent that modifies the parameter $\zl$. This is done many times, repeating the same protocol 
$\lambda:[0,\tau]\rightarrow\mathbb{R}$ over a given finite time $\zt$, changing the parameter from its initial value $\lambda(0)=A$, to its final value $\lambda(\zt)=B$. Each time, a different initial condition is taken at random, according to the canonical distribution \eqref{CanEns}, and the following quantity, called work, is computed \cite{jarzynski1997nonequilibrium}:
    \be
    W_J(\Gamma_0) = \int_0^\tau \, {\partial H \over \partial \lambda} \, \dot\lambda \,  \mbox{d}  t =
{\cal H}(\Gamma_\zt(\Gamma_0); B) -
{\cal H}(\Gamma_0; A) 
    \label{WJwork}
\ee
where $\Gamma_\zt(\Gamma_0)$
is the phase reached in the time $\zt$ starting from the initial condition $\Gamma_0$.
Because the protocol $\lambda(t)$ is fixed, the dynamics are deterministic, and the value of the work depends only on the initial condition. However,
the initial conditions change randomly, yielding a different value of $W_J$ for each realization of the process, and effectively making it a random variable. In this setting, the following relation, known as Jarzynski equality, was obtained: 
\cite{jarzynski1997nonequilibrium}:
    \be
    \Big\langle e^{-\beta W_J} \Big\rangle_0 = e^{-\beta \Delta F}
    \label{JE}
    \ee
where $\langle \cdot \rangle_0$ is the canonical ensemble average obtained from $P_0$, and
$\Delta F=F_B-F_A$ is the equilibrium free energy difference between the equilibrium canonical state with parameter $\lambda=B$ and the one with parameter $\lambda=A$, both at temperature $T$. 
One of the most striking aspects of the JE, that is a direct effect of the canonical ensemble, is that it does not depend on the protocol. This sounds at odds with the fact that physical theories have a range of applicability limited by space and time constraints, outside of which a different description must be adopted. On the other hand, it depends on the validity of the canonical ensemble whose applicability boundaries are not 
known, in general, especially if involving non standard quantities. Understanding the role of the canonical ensemble is important in general, not just in relation to the JE. 
We will see that the quantity in the right hand side of  Eq.\eqref{JE}, depends on the protocol, if the ensemble does not extend to infinity. Note that the form of the probability distribution properly describing the effect of finite environments are not known in general, but the finite size effects can be evidenced on any distribution. In the concluding remarks we address this issue.

\section{Models and methods}
	\label{s:models_methods}
    \par\noindent
Below, we investigate possible finite size effects for several different systems. In particular, we analyze three simple  harmonic oscillators models, perturbed from their equilibrium states. The perturbation is applied by harmonic springs, whose center of force moves according to  deterministic rules $\zl(t)$. The first model consists of a single oscillator, playing the role of S, with $\lambda(t)=\ell t$, $t \in [0,\tau]$, where $\ell$ and $\zt$ are constants that can be varied, in such a way that the initial and final values of $\zl$ do not change: $\zl(0)=A$ and $\zl(\zt)=B$.
In the second model, the protocol is changed to $\lambda(t)=\sin(\gamma t)$. Being periodic in time, this protocol yields  different phenomena when the frequency $\gamma$ is changed, like resonances that affect $W_J$ and, consequently the JE.
Both, the first and the second case do not have any environment E or, equivalently, the interaction energy vanishes: $h_{I}=0$.  
The third model we consider has two oscillators, one of which is taken to be the system S and the other the environment E. As the theory requires, only S is subjected to a time dependent perturbation.

\subsection{Single oscillator under linear protocol}
	\label{ss:linear_forcing}
	\par\noindent
Take a 1D system made of a single harmonic oscillator with rest position in $x=0$, that is driven by a moving harmonic trap, centered in $\lambda(t) = \ell t$, where $\ell$ is a positive constant, and $t \in [0,\tau]$. The initial value of $\zl$ is given by $A=\lambda(0)=0$, and let its final value be denoted by $B=\lambda(\tau)=\ell \tau$, with $B$ fixed. To explore the effect of modifying the speed of the protocol, we vary $\ell$ and $\tau$, so that $B$ is fixed. Let the oscillator mass be $m$, and its momentum $p=mv$, where $v$ is the velocity. Then, the motion is determined by the following time dependent Hamiltonian:
	\bea
	{\cal H}(x,v;t) &=& {p^2 \over 2 m} + {k_p \over 2} x^2 + {k_D \over 2} \left(\lambda - x \right)^2 
	= {p^2 \over 2m} + {k \over 2} x^2 + {k_D \ell \over 2} \left(\ell t^2 - 2 x t \right)~,
	\label{ham}
	\eea
where $k_p$ is the elastic constant of the spring with rest position in $x=0$, $k_D$ the elastic constant of the moving trap, and $k = k_D + k_p$. 
The equation of motion consequently takes the form:
	\be
	\ddot x = - \omega^2 x + {k_D \over m} \ell t ~, 
	\quad \mbox{with } ~ x(0) = x_0 \, , ~ v(0) = v_0 
	\label{eqmot}
	\ee
where we introduced the natural frequency of the oscillator $\omega=\sqrt{k/m} $. In this case, the work $W_J$ is expressed by:
	\bea
	&&W_J = \int_0^\tau \, k_D \left( \ell t - x \right) \ell \,  \mbox{d}  t = {k_D \ell^2 \tau^2 \over 2} - 
	k_D \ell \int_0^\tau \, x(t;x_0,v_0) \, \mbox{d}  t \nonumber \\
	&& \hskip 20pt = k_D B \left[ {B \over 2} - {1 \over \tau} \int_0^\tau \, x(t;x_0,v_0) \, \mbox{d}  t \right]
	\label{WJ}
	\eea
where the oscillator position is expressed by:
	\be
	x(t;x_0,v_0) = x_0 \cos \omega t + {v_0 - \ell k_D/k \over \omega} \sin \omega t + {\ell k_D \over k} t
	\label{solmot}
	\ee
Then, performing the integration in expression \eqref{WJ}, one obtains:
	\be
	W_J(\ell; x_0 , v_0) = k_D B \left[ {B \over 2} \left(1 - {k_D \over k} \right) - {x_0 \ell \over B \omega} 
	\sin \omega  {B \over \ell} + 
	\left({p_0 \ell \over B k} - {\ell^2 k_D \over B k \zw^2} \right) \left( \cos \omega  {B \over \ell} - 1 \right)
	\right] ~,
	\label{WJexpl}
	\ee
where $B$ is fixed, while the protocol speed $\ell$ can be varied. Although $\exp(-\beta W_J)$ depends on $\ell$, its average with respect to the initial canonical ensemble, $P_0$, does not. 	
Given $\ell \in (0,\infty)$, one has:

\be
\Big\langle e^{-\beta W_{J,\ell}} \Big\rangle_0 = \exp \left\{ -\beta {k_D k_p B^2 \over 2 k} \right\} \; ,
	\label{aveWJlfin}
 \ee
which does not depend on the speed of the protocol, as the Jarzynski theory predicts.
Explicit calculations are reported in the Appendix \ref{app:app1}.

In the case in which the environment is bounded and the bath can only express a finite energy, the corresponding probability density is truncated at a given distance $L$ from the rest position of the oscillator, and at a maximum momentum $M$. 
For the sake of argument, we assume that the form of the finite support distribution is the canonical one, truncated and normalized, and that the two bounds $L$ and $M$ do not depend on each other. After all, the classical ensembles constitute a most successful postulate of statistical mechanics that, however, only seldom can be derived from the particles dynamics. Morevoer the resulting distributions are truncated Gaussians, hence mitigte the effects of truncation. Then, suppose we have:
\be
P_0(x,p) = \cfrac{1}{Z_0(L,M)}
\left\{
\begin{array}{lcr}
 e^{-\beta {(k x^2 + p^2/m)/ 2}} & \mbox{if} & | x | \le L ~\mbox{ and }~ |p| \le M \\[7pt]
0  & \mbox{if} & | x | > L ~\mbox{ or }~ |p| > M  
\end{array}
\right.
\label{P0xp}
\ee
with $Z_0(L,M)$ a normalizing factor.
	In this case, one obtains:
	\be
\Big\langle e^{-\beta W_{J,\ell}} \Big\rangle_{0;L,M} = %\cfrac{1}{Z_{0;L,M}}\cdot 
 I_{exp}\cdot I_x \cdot I_p
 \label{eq:3I}
	\ee
%	with
%	\bea
%	&&I_{exp} = \exp \left\{-\beta {k_D k_p B^2 \over 2 k } \right\} = 
% \Big\langle e^{-\beta W_{J,\ell}} \Big\rangle_0 \\[5pt]
%	&&I_x = { \mbox{erf} \left( \sqrt{\beta k \over 2} L + \sqrt{\beta m \over 2} 
%		{k_D  \over k}\ell \sin \omega {B \over \ell}  \right) 
%		+ \mbox{erf} \left( \sqrt{\beta k \over 2} L - \sqrt{\beta m \over 2} 
%		{k_D  \over k}\ell \sin \omega {B \over \ell}
%		\right) \over 2 ~ \mbox{erf}\left(\sqrt{\beta k \over 2} L \right) } \\[5pt]
%	&&I_p = { \mbox{erf} \left( \sqrt{\beta \over 2 m} M + \sqrt{\beta m \over 2} {k_D\over k} \ell  
%		\left( \cos \omega  {B \over \ell} - 1 \right)
%		\right) 
%		+ \, \mbox{erf} \left( \sqrt{\beta \over 2 m} M - \sqrt{\beta m \over 2} {k_D \over k} \ell 
%		\left( \cos \omega  {B \over \ell} - 1 \right)
%		\right) \over 2 ~ \mbox{erf}\left(\sqrt{\beta \over 2 m} M \right) } \nonumber
%	\eea
where $I_{exp}$ represents the infinite size result, that does not depend on $\ell$, while the finite size correction factors $I_x$ and $I_p$ do depend on $\ell$, hence on the protocol. The explicit expressions of $I_{exp},I_x,I_p$, along with the detailed calculations leading to Eq. \eqref{eq:3I}, are deferred to the Appendix \ref{app:app1}.
This result shows that for fixed $\ell$ and $m$, sufficiently large $L$ and $M$ exist such that the infinite size result is recovered; indeed $I_x$ and $I_p$ both tend to $1$, if $L,M$ grow at fixed $\ell$.
However, for fixed $L$ and $M$,  sufficiently large $\ell$, {\em i.e.}\ a sufficiently fast protocol, together with a large enough value of the product $\zw B$, or sufficiently large $m$, yield $I_x , I_p < 1$, {\em i.e.}\ $\exp(-\beta W_J)\rangle_{0;L,M} < \langle \exp(-\beta W_J)\rangle_0$. 
The term $I_p$ is particularly sensitive to variations of $m$, because the argument of the error function on the right of its numerator may even turn negative, if $m$ is sufficiently large. In any event, the left hand side of the JE is protocol dependent, if the ensemble is finitely supported. While at odds with the infinite bath result, this is in accord with the fact that too fast protocols ({\em e.g.}\ comparable with microscopic rates) require a specifically developed approach.

%This comes as a direct consequence of the properties of the error functions appearing in $I_x$ and $I_p$, resulting from the restriction of the integration domains in the computation of averages.
	
	\subsection{Single Oscillator with Periodic Forcing}
	\label{ss:periodic_forcing}
	\par\noindent
	
Let now the single oscillator  be driven by a moving harmonic trap centered in $\lambda(t) = \sin{\gamma t}$, where $\gamma={2\pi}/{T}$, and $T$ is the period of the center of force of the trap. Take $t \in [0,\tau]$, $A=\lambda(0)=0$, and  $B=\lambda(\tau)=\sin{\gamma \tau}$.
If the final value of $\zl$ is fixed, as in the previous subsection, different $\gamma$ correspond to faster or slower protocols, that last a time $\tau = \arcsin({B})/\gamma$. The time dependent Hamiltonian now takes the form
	\bea
	{\cal H}(x,v;t) &=& {p^2 \over 2 m} + {k_p \over 2} x^2 + {k_D \over 2} \left(\lambda - x \right)^2 \\
	&=& {p^2 \over 2m} + {k_p \over 2} x^2 + {k_D \over 2} \left(\sin{\gamma t} - x \right)^2
	= {p^2 \over 2m} + {k \over 2} x^2 + {k_D \over 2} \left( \sin^2 \gamma t - 2 x \sin{\gamma t} \right)~,
	\label{ham2}
	\eea
	where $k = k_D + k_p$.
	In this case, the Jarzynski work $W_J$ is given by:
	\bea
	&&W_J = \int_0^\tau \, {\partial H \over \partial \lambda} \, \dot\lambda \,  \mbox{d}  t = \\
%	\int_0^\tau \, k_D \gamma \left( \sin \gamma t - x(t;x_0,v_0) \right) \cos \gamma t \,  \mbox{d}  t = \\
%	&& \hskip 20pt = k_D \gamma \int_0^\tau \, \sin \gamma t \cos \gamma t  \, \mbox{d}  t - 
%	k_D \gamma \int_0^\tau \, x(t;x_0,v_0) \cos \left( \gamma t \right) \, \mbox{d}  t \\
%	&& \hskip 20pt = {k_D \gamma \over 2} \int_0^\tau \, \sin 2 \gamma t  \, \mbox{d}  t - 
%	k_D \gamma \int_0^\tau \, x(t;x_0,v_0) \cos \left( \gamma t \right) \, \mbox{d}  t = \\
	&& \hskip 20pt =  {k_D \over 4} \left( 1 - \cos 2\gamma \tau \right) - k_D \gamma \int_0^\tau x(t;x_0,v_0) \cos \left( \gamma t \right) \, \mbox{d}  t
	\label{WJ2}
	\eea
	Given the Hamiltonian \eqref{ham2}, the equation of motion for this system is:
	\bea
	&& 
	\ddot x = - \zw^2 x + {k_D \over m} \sin \gamma t ~, 
	\quad \mbox{with i.c.\ } ~ x(0) = x_0 \, , ~ v(0) = v_0
	\label{eqmot2}
	\eea
where $\zw = \sqrt{k/m}$. For $\gamma\neq\omega$, one obtains:
	\bea
	&& x(t;x_0,v_0) = {k_D/m \over \omega^2 - \gamma^2} \sin \gamma t + {1 \over \omega}\left( v_0 - {\gamma k_D/m\over \omega^2 - \gamma^2} \right) \sin \omega t + x_0 \cos \omega t
	\label{lawmot2}
	\eea
and the work takes the form:
	\bea
	&&\hskip -60pt W_J(\gamma;x_0,v_0) = {k_D \over 4} \left( 1 - \cos 2\gamma \tau \right) -k_D \gamma \int_0^\tau \cos \left( \gamma t \right) \times \\[5pt] 
	&& \hskip 30pt \left[ {k_D/m \over \omega^2 - \gamma^2} \sin \gamma t + {1 \over \omega}\left( v_0 - {\gamma k_D/m\over \omega^2 - \gamma^2} \right) \sin \omega t + x_0 \cos \omega t \right]  \, \mbox{d}  t
	\label{WJ3}
	\eea
	Solving the integral on the right, we finally get:
	\bea
	&&\hskip -50pt
 W_J(\gamma;x_0,v_0) = {k_D \over 4}\left( 1-{k_D/m \over \omega^2-\gamma^2}\right) \left( 1 - \cos 2\gamma \tau \right) + \\[5pt] 
	&& \hskip 20pt -{k_D\gamma\over\omega^2-\gamma^2}\left(v_0 -{k_D\gamma/m\over\omega^2-\gamma^2}\right)\left(1-{\gamma\over\omega}\sin\gamma\tau\sin\omega\tau-\cos\gamma\tau\cos\omega\tau\right) + \\[5pt] 
	&& \hskip 20pt + \, x_0 {k_D\gamma\omega\over\omega^2-\gamma^2}\left( {\gamma\over\omega}\sin\gamma\tau\cos\omega\tau - \cos\gamma\tau\sin\omega\tau \right)
	\label{WJfin}
	\eea
This quantity can now be multiplied by $-\beta$, exponentiated and averaged over all the initial conditions $(x_0,v_0)$.
In the case of the full canonical ensemble, one obtains a result that does not depend on $\gamma$, when $A$, $\zt$ and consequently $B$ are fixed. 
If, on the other hand, the probability density is expressed by Eq.\eqref{P0xp},
one finds:	
%	\bea
%	&&\hskip -40pt \Big\langle e^{-\beta W_J} \Big\rangle_{0;L,M} = { 1 \over Z_0(L,M)} \times  \nonumber \\[5pt]
% &&\hskip -25pt 
% \exp\left\{ -{\beta k_D \over 4}\left( 1-{k_D/m \over \omega^2-\gamma^2}\right) \left( 1 - \cos 2\gamma \tau \right) - {\beta k_D^2\gamma^2/m \over (\omega^2-\gamma^2)^2}\left(1-{\gamma\over\omega}\sin\gamma\tau\sin\omega\tau  -\cos\gamma\tau\cos\omega\tau\right) \right\} \times \nonumber \\[5pt]
	%\vphantom{\frac12}\right]
%	&& \int_{-L}^L \mbox{d} x \, 
% \exp\left\{-{\beta k \over 2}x^2 + {\beta k_D\gamma\omega \over \omega^2-\gamma^2} \left( \cos\gamma\tau\sin\omega\tau -  {\gamma\over\omega}\sin\gamma\tau\cos\omega\tau\right)x \right\} \times 
% \nonumber \\[5pt]
%	&& \int_{-M}^M \mbox{d} p \, 
% \exp\left\{- { \beta \over 2 m}p^2 + {\beta k_D\gamma/m\over\omega^2-\gamma^2}\left(1-{\gamma\over\omega}\sin\gamma\tau\sin\omega\tau-\cos\gamma\tau\cos\omega\tau\right)p
%  \right\}
%	\label{av_exp_WJ_expanded}
%	\eea
%where the partition function $Z_0(L,M)$ is the same as the one given in \eqref{FinSiz}. Computing the integrals, 
%one may eventually write:
	\be
	\Big\langle e^{-\beta W_J} \Big\rangle_{0;L,M} = I_{exp} \cdot I_x \cdot I_p  \; ,
	\label{av_exp_WJ_explicit}
	\ee
where the explicit expressions of $I_{exp},I_x,I_p$ are given in the Appendix \ref{app:app2}.
The \textit{resonance}, corresponding to $\gamma=\omega$, must be treated separately, since the solution of the equation of motion  \eqref{eqmot2} takes the form: 
	\bea
	&& x(t;x_0,v_0) = x_0 \cos \omega t + {v_0\over \omega} \sin \omega t + {k_D/m\over 2 \omega^2}\left(\sin\omega t -\omega t \cos \omega t \right)
	\label{lawmot2_resonance}
	\eea
The Jarzynski work is now expressed by:
%	\bea
%	&&\hskip -40pt 
% W_J = {k_D \over 4} \left( 1 - \cos 2\omega \tau \right) \nonumber \\
%&& -k_D \omega \int_0^\tau \cos \left( \omega t \right)  \left[ x_0 \cos \omega t + {p_0\over \omega m} \sin \omega t + {k_D/m\over 2 \omega^2}\left(\sin\omega t -\omega t \cos \omega t\right) \right]  \, \mbox{d}  t
%	\label{WJ_resonance}
%	\eea
%	which yields
	\bea
	&&\hskip -40pt
 W_J(\zt;x_0,v_0) = {k_D^2/m \over 8}\tau^2 + {k_D^2/m \over 8\omega}\tau\sin 2\omega\tau + {k_D\over 4}\left(1-{3\over 4}{k_D/m\over\omega^2}\right)\left[1-\cos 2\omega\tau\right] \nonumber \\
	&& \hskip 60pt - x_0{k_D\over 2}\left[\omega\tau+{1\over 2}\sin 2\omega\tau\right]-v_0{k_D\over 4\omega}\left[1-\cos 2\omega\tau\right]
	\label{WJfin_resonance}
	\eea
and its finite energy ensemble average can again be written as:
	\be
	\Big\langle e^{-\beta W_J} \Big\rangle_{0;L,M}^{\rm (res)} = I_{exp}^{\rm (res)} \cdot I_x^{\rm (res)}  \cdot I_p^{\rm (res)}
	\label{av_exp_WJ_explicit_resonance}
	\ee
	where:
	\bea
	&& I_{exp}^{\rm (res)}=\exp \left\lbrace {\beta k_D\over 2}\left( {k_D\over k}-1 \right)\sin^2\omega\tau \right\rbrace
	\label{Iexp_resonance}
	\eea
	
	\bea
	&& I_x^{\rm (res)}= {1\over 2 ~ \mbox{erf}\left(\sqrt{\beta k \over 2 } L \right) } \left[ \mbox{erf}\left({\beta k L - {\beta k_D \over 2}\left(\omega\tau+{1\over 2}\sin2\omega\tau\right)\over \sqrt{2\beta k} } \right) + \right. \nonumber \\
	&& \hskip 110pt \left.\mbox{erf}\left({\beta k L + {\beta k_D \over 2}\left(\omega\tau+{1\over 2}\sin2\omega\tau\right)\over \sqrt{2\beta k} } \right) \right]
	\label{Ix_resonance}
	\eea
	
	\bea
	&& I_p^{\rm (res)} = {1\over 2 ~ \mbox{erf}\left(\sqrt{\beta  \over 2 m} M \right) } \left[ \mbox{erf}\left({{\beta\over m} M - {\beta k_D /m\over 4\omega}\left(1-\cos2\omega\tau\right)\over \sqrt{2\beta /m} } \right) + \right. \nonumber \\
	&& \hskip 110pt \left. \mbox{erf}\left({{\beta\over m} M + {\beta k_D /m\over 4\omega}\left(1-\cos2\omega\tau\right)\over \sqrt{2\beta /m} } \right) \right]
	\label{Ip_resonance}
	\eea
Because $\zw$ can be considered an intrinsic property of the system coupled to the driving mechanism, we take it as fixed. Then, Equation \eqref{av_exp_WJ_explicit_resonance}, together with \eqref{Iexp_resonance}-\eqref{Ip_resonance}, shows that the average of the exponential of the Jarzynski work for a bounded ensemble of initial states depends on the protocol time $\tau$. Indeed, for a sinusoidal protocol, there is an infinite set of values of $\zt$ that yields the same final value $\zl(\zt)=B$. In particular,
Eq.\eqref{Ix_resonance}, shows that 
$I_x^{\rm (res)}$ may even approach $0$ or $2$, however large $L$ is taken, for sufficiently large $\zt$. Indeed, the first error function in Eq.\eqref{Ix_resonance} tends to $-1$, while the other tends to 1, as $\zt$ grows, all the other parameters being fixed. On the other hand, small $\zt$ implies a sum of two equal quantities, which approaches 2 for large $L$. Tuning the values of $\zt$ one observes quite a sensitive protocol dependence, for the average \eqref{av_exp_WJ_explicit_resonance}.
This is illustrated in Figs.\ \ref{fig:avexpWj_varB} and \ref{fig:avexpWj_varK}. The cause of this behaviour, in presence of a resonance, is the fact that the amplitude of the oscillator position grows linearly in time, yielding the  $\omega\tau$ term in the arguments of the error functions of $I_x^{\rm (res)}$.

\subsection{Coupled Oscillators with Periodic Forcing}
\label{pairOS}
%	\label{ss:coup_oscill_periodic_forcing}
	\par\noindent
In this subsection, a single oscillator, $S$, is  harmonically tied to the origin of the real line, and is  harmonically driven, as in Sec.\ \ref{ss:periodic_forcing}. In addition, S is harmonically coupled to a second oscillator, E. We denote by $k_I$ the stiffness of the harmonic potential linking S and E, and we also assume that
E is harmonically bound to the origin of the line, with elastic constant $k_E$. Let the oscillators masses be $m_E$ and $m_S$, and let the phase  of S+E be denoted by $\Gamma =\left(x_E,x_S,p_E,p_S\right) = \left(\mathbf{x},\mathbf{v}\right)$, where $p_E=m_Ev_E$ and $p_S=m_Sv_S$ are the momenta associated to each oscillator. Then, the Hamiltonian of the total system is given by:
	\bea
	{\cal H}(\mathbf{x},\mathbf{v};\zl) &=&
 \left[ 
 {p_S^2 \over 2 m_S} + {k_S \over 2} x_S^2 + {k_D \over 2} \left(\lambda - x_S \right)^2 \right] +
 \left[ {p_E^2 \over 2 m_E} + {k_E \over 2} x_E^2 \right] + {k_I \over 2} (x_E - x_S) ^2
 \\[5pt]
	&=& H_S(x_S,p_S; \zl) + H_E(x_E,p_E) +h_I(\mathbf{x})
\label{eq:ham_coup_oscill}
	\eea
where the square brackets delimit the different contributions to the full Hamiltonian, respectively $H_S ,\; H_E$ and $h_I$, as in Eq.\eqref{JEham} for the JE theory. 
As driving term, we take the periodic protocol used above: $\sin \gamma t$, and we set again  $k=k_S+k_D$.
The equations of motion  for this system are the following:
\bea
	\begin{cases}

	m_S \ddot x_S = - k x_S + k_D \sin \gamma t - k_I \left(x_S-x_E\right)~,\\ 

	m_E \ddot x_E = - k_E x_E - k_I \left(x_E-x_S\right)~,
	\end{cases}
	\label{eqmot_couposcill}
	\eea
 with initial conditions $\left(\mathbf{x}(0),\mathbf{\dot x} (0)\right) = \left(\mathbf{x}_0,\mathbf{v}_0\right) = \Gamma_0$.
While analytical solutions for this set of equations are conceptually trivial, they are practically involved if $k_S\neq k_E$ and $m_S\neq m_E$, especially when integrated to compute the left hand side of the JE. On the other hand, they can be quite simply handled in numerical calculations.
We have thus numerically sampled the initial conditions $\Gamma_0$ from the truncated canonical distribution, and for each of them we have computed the initial energy ${\cal H}(\Gamma_0;\zl(0))$. Then, we have numerically solved Eqs.\ 
\eqref{eqmot_couposcill} for that $\Gamma_0$, obtaining the final condition $\Gamma_\zt(\Gamma_0)$, that has  been introduced in the final Hamiltonian ${\cal H}(\Gamma_\zt(\Gamma_0);\zl(\zt))$, to obtain the work as:
\be
W_J ( \Gamma_0) = {\cal H}(\Gamma_\zt(\Gamma_0);\zl(\zt)) - {\cal H}(\Gamma_0;\zl(0))
\ee
as in Eq.\ \eqref{WJwork}, where $\zl(0)=A=0$ and $\zl(\zt) = \sin \gamma \zt = B$. Collecting many works, with $\zt$ fixed, we have eventually estimated
the quantity 
\be
\Big\langle e^{-\zb W_J}
\Big\rangle_{0;L,M} 
\label{LM0}
\ee

%The particular case where $m_E\gg m_S\approx 1$ and $k_E\sim m_E$, $k_I\sim m_S$, which is of interest when one wants to consider an environment with much larger inertia then the system, is analitically approachable since the system \eqref{eqmot_couposcill} can be reduced, in the limit, to
%\bea
%	\begin{cases}

%	\ddot x_S = - \frac{k}{m_S} x_S + \frac{k_D}{m_s} \sin \gamma t - \frac{k_I}{m_S} \left(x_S-x_E\right)~,\\ 

%	\ddot x_E = - \omega_E^2 x_E ~,
%	\end{cases}
%	\label{eqmot_couposcill_reduced}
%\eea
%again with initial conditions $\left(\mathbf{x}(0),\mathbf{\dot x} (0)\right) = \left(\mathbf{x}_0,\mathbf{v}_0\right) = \Gamma_0$ and where $\omega_E^2=k_E/m_E$.
%The two equations are thus decoupled, with the second one being the equation of motion of a simple harmonic oscillator with no external forcing. Assuming for simplicity that $\dot x_E(0)=0$, and introducing the law of motion for the environment oscillator in the first equation, leads to
%\be
%	\ddot x_S = - \omega_S^2 x_S + \frac{k_D}{m_s} \sin \gamma t + \frac{k_Ix_{E,0}}{m_S} \cos\omega_E t
%	\label{eqmot_decouposcill}
%\ee
%where $\omega_S^2=(k+k_I)/m_S$. Its explicit solution is given by
%\bea
%	x_S(t) = && \frac{1}{m_S}\left[\frac{k_D}{\omega_S^2-\gamma^2}\sin\gamma t + \frac{k_Ix_{E,0}}{\omega_S^2-\omega_E^2}\cos\omega_E t\right] +\\
%	&& \left( x_{S,0}-\frac{k_Ix_{E,0}}{m_S\left(\omega_S^2-\omega_E^2\right)}\right)\cos\omega_St + \frac{1}{\omega_S}\left(v_{S,0}-\frac{\gamma k_D}{m_S\left(\omega_S^2-\gamma^2\right)}\right)\sin\omega_St
%	\label{eqmot_xS_couposcill}
%\eea

\subsection{Results}
    \label{s:results}
    \par
\begin{figure}[!ht]
    	\centering
    	\subfigure{\includegraphics[width=0.40\textwidth]{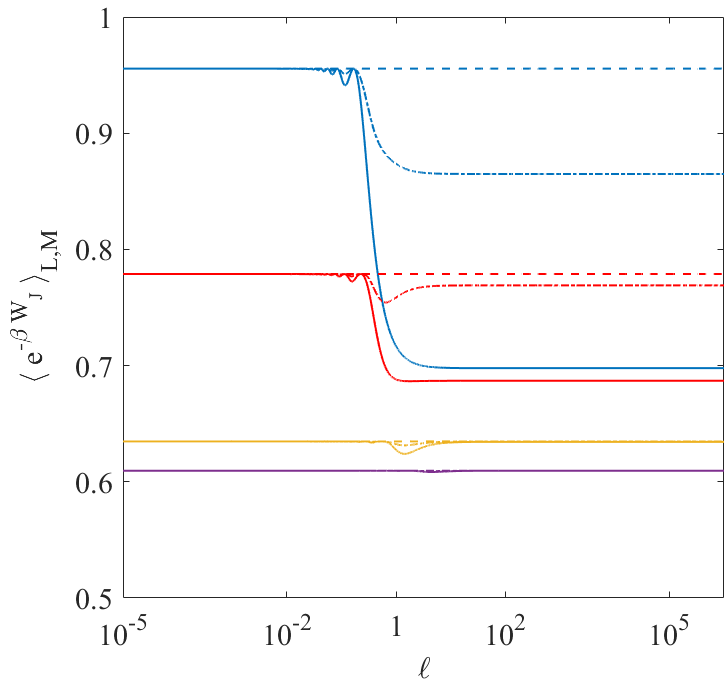}}\hspace{0.9cm}
    	\subfigure{\includegraphics[width=0.40\textwidth]{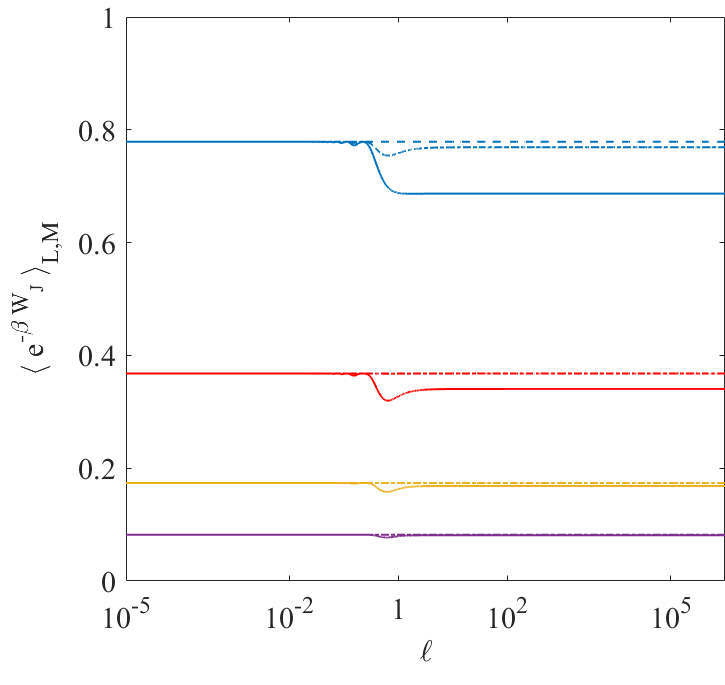}}
    	\caption{Values of $\ensavg_{L,M}$ for the single harmonic oscillator driven by a constant speed moving harmonic trap, with final protocol value of $\lambda(\tau)=B=1$. The result is shown as a function of $\ell$, for different values of $L$ and $M$. Solid lines refer to $L=M=1$, dash-dotted lines to $L=M=2$, and dashed lines to $L=M=5$. In the left panel, blue, red, yellow and purple plots refer to $k_p=0.1,\;1,\;10,\;100$, respectively. In the right panel, blue, red, yellow and purple plots refer to $\beta=1,\;4,\;7,\;10$, respectively. Other parameters are set to $m=1$ and $k_D=1$.}
    	\label{fig:avexpWj_varkp}
    \end{figure}

\noindent
Our first observation is that finite size effects make the protocol dependent the quantity \eqref{LM0} unlike the case of systems initially in contact with truly infinite reservoirs. 
Of course, no real reservoir is infinite, but considering it infinite introduces no errors when taking equilibrium averages of standard observables, such as power laws. The situation changes if exponentials of standard observables are considered.
For the single oscillator driven by a harmonic trap moving with constant velocity, Fig. \ref{fig:avexpWj_varkp} shows  the dependence of  \eqref{LM0} on $\ell$ and on $\beta£$, for different values of the harmonic potential stiffness $k_p$:
fast and slow protocols yield different ensemble averages. The cases with $L=1$ and $M=1$, represented by solid lines, show an abrupt transition at about $\ell=1$, for small $k_p$. For large $k_p$, dominating the coupling with the driving agent, the result gradually turns independent of the speed of the process. 
Increasing the reservoir size to $L=5$ and $M=5$, the quantity \eqref{LM0} does not appear to depend anymore on the speed of the process  $\lambda(t)$, cf.\ dashed lines in Fig.\ \ref{fig:avexpWj_varkp}. In reality, the dash-dotted lines for 
$L=M=2$ reveal that the process dependence merely shifts with $L$ and $M$, becoming evident at larger $\ell$. Therefore, process independence for \eqref{LM0} is only obtained when $L=\infty$ and $M=\infty$. An analogous behaviour is observed as a function of the inverse temperature $\beta$, with more evident transitions at higher temperatures.
    \FloatBarrier

\begin{figure}[!ht]
\centering
\subfigure{\includegraphics[width=0.40\textwidth]{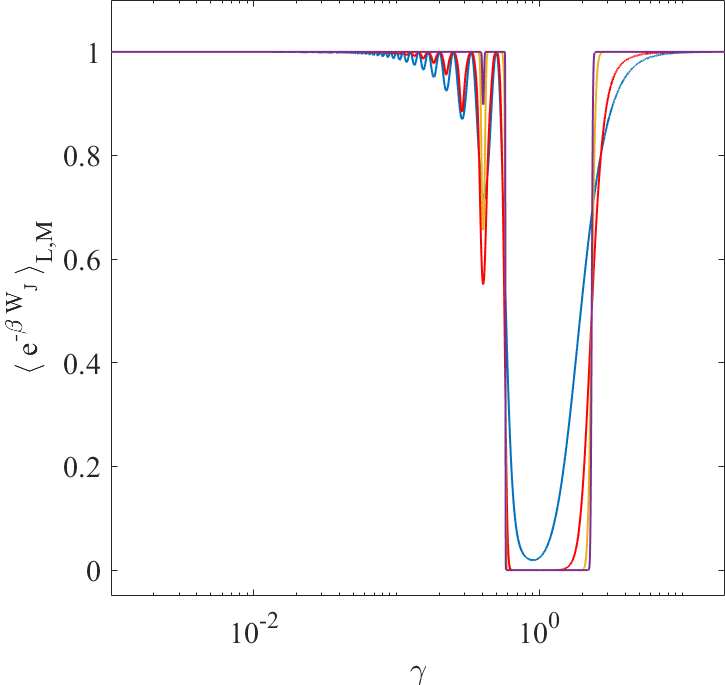}}\hspace{0.9cm}
\subfigure{\includegraphics[width=0.40\textwidth]{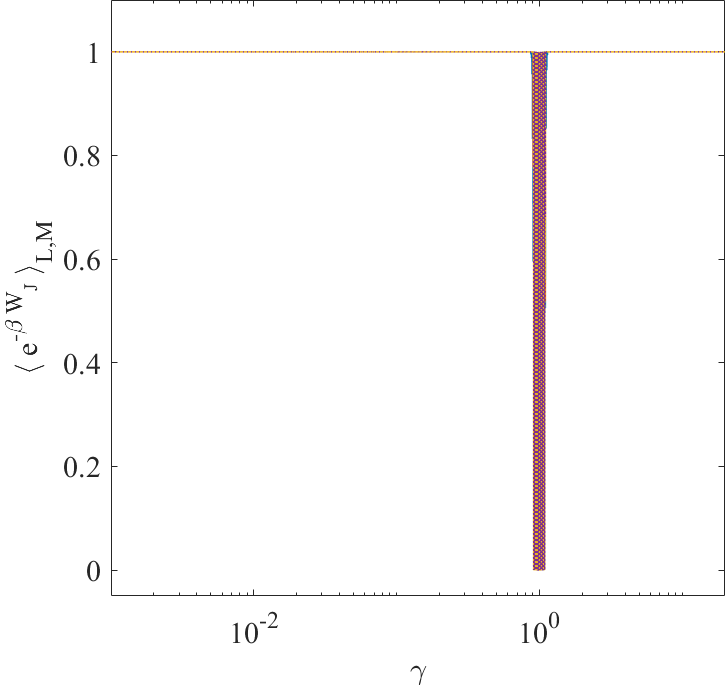}}		%\hfill
\caption{Values of $\ensavg_{L,M}$ for the single harmonic oscillator with $\lambda=\sin\gamma t$, as a function of the forcing frequency $\gamma$, for different values of $\Gamma_0$ volumes and different final times $\tau$. Left and right panels refer to $L=M=1$ and $L=M=10$ respectively, with $\tau$ such that $B=\sin(2\pi)$ for the first case, and $B=\sin(200\pi)$ for the second. Blue, red, yellow and purple plots refer to inverse temperatures $\beta=1,\;10,\;100,\;1000$, respectively. Other parameters are set to $m=1$, $k=1$, $k_D=1$.}
\label{fig:avexpWj_varB}
\end{figure}

\begin{figure}[!ht]
    	\centering
    	\subfigure{\includegraphics[width=0.40\textwidth]{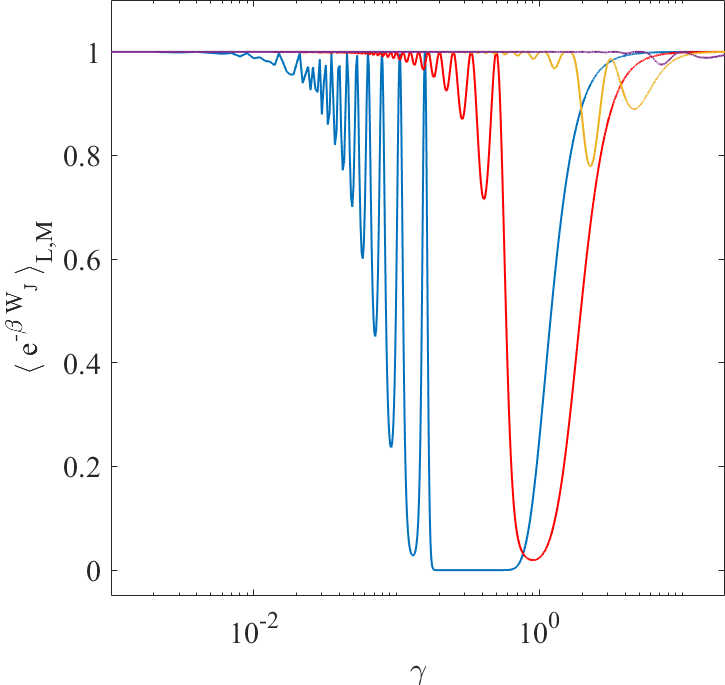}}\hspace{0.9cm}
    	\subfigure{\includegraphics[width=0.40\textwidth]{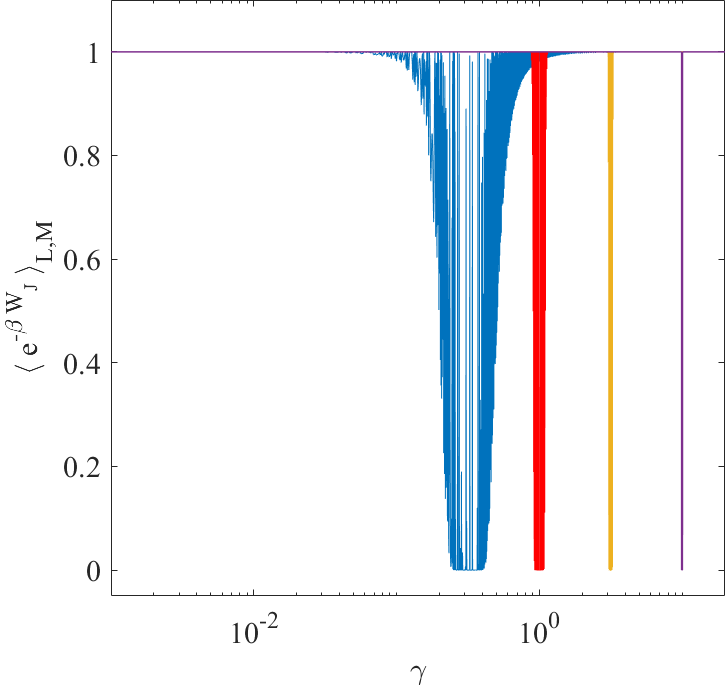}}		%\hfill
    	\caption{Values of $\ensavg_{L,M}$ for the single harmonic oscillator with $\lambda=\sin\gamma t$, as a function of the forcing frequency $\gamma$, for different values of $\Gamma_0$ volumes and different final times $\tau$. Left and right panels refer to $L=M=1$ and $L=M=10$ respectively, with $\tau$ such that $B=\sin(2\pi)$ for the first case and $B=\sin(200\pi)$ for the second. Blue, red, yellow and purple plots refer to stiffnesses $k_s=0.1,\;1,\;10,\;100$, respectively. Other parameters are set to $m=1$, $\beta=1$ and $k_D=1$.}
    	\label{fig:avexpWj_varK}
\end{figure}

The second model analyzed above is even more intriguing, as resonances significantly affect the work done on the system by external perturbations,  when finite size effects play a role. 
Figures \ref{fig:avexpWj_varB} and \ref{fig:avexpWj_varK} show that the extension of the phase space volume does not suffice to tame the resonances produced over sufficiently long times $\tau$. Unlike the case of infinitely large baths, which yield the same result for all $\tau$, here a  protocol dependence arises. The reason is that
a harmonic oscillator subject to no friction and to periodic forces performs oscillations whose amplitude grows linearly in time, if the forcing frequency equates the natural frequency of the system. In our example, this happens for $\gamma=\zw$. Thus, the work done on the system grows together with the amplitude, pushing $\ensavg_{L,M}$ toward $0$, at and near the resonance. 

\FloatBarrier
The inverse temperature $\beta$ and the global stiffness $k$ of the potential also have noticeable effects on the behavior of \eqref{LM0}. In particular, increasing $\beta$ (reducing the temperature) flattens the curve about the resonance, widening the interval of $\gamma$ values that make $\ensavg_{0;L,M}$ equal 0, rather than the infinite size theoretical value 1, cf.\ Fig. \ref{fig:avexpWj_varB}.
A larger value of $k$ seems instead to stabilize $\ensavg_{0;L,M}$ and reduce its dependence on $\gamma$, as shown by Fig.\ \ref{fig:avexpWj_varK}.     
\begin{figure}[!ht]
\centering
\subfigure{\includegraphics[width=0.30\textwidth]{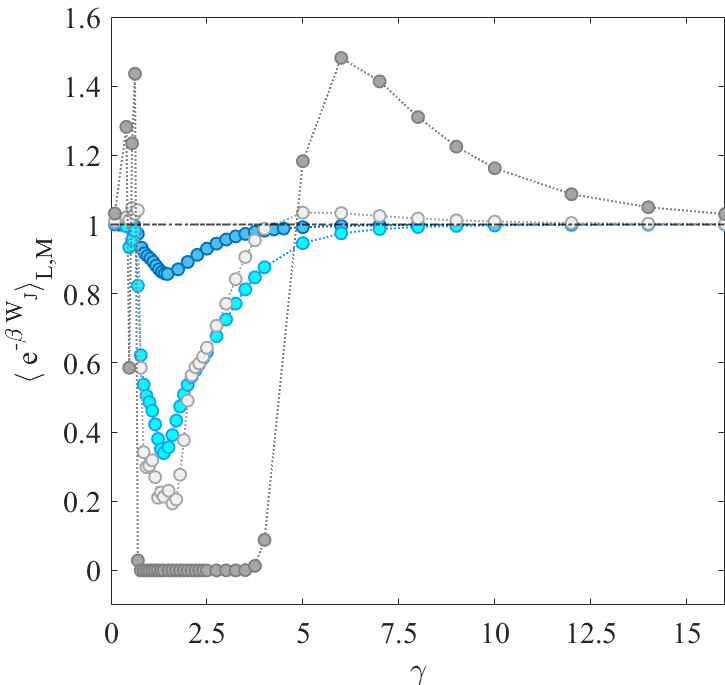}} \hspace{0.5cm}
\subfigure{\includegraphics[width=0.30\textwidth]{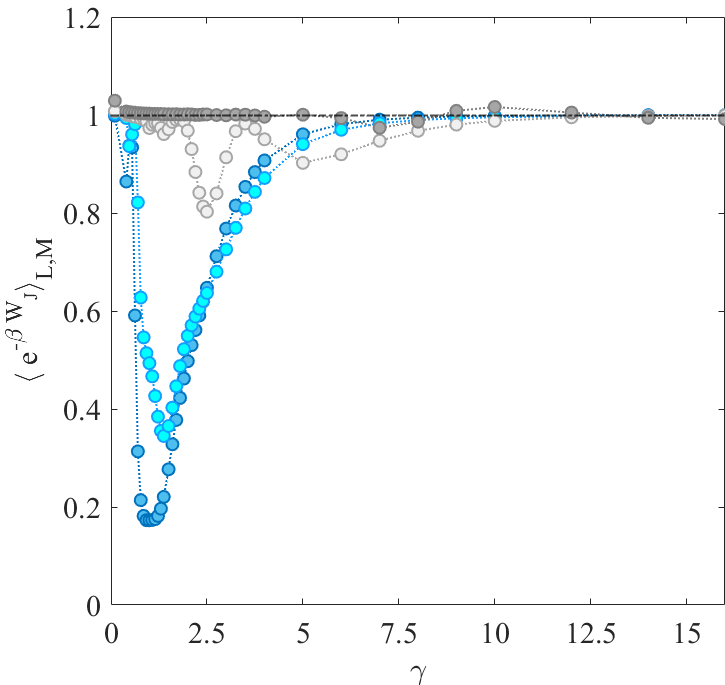}}	\hspace{0.5cm}
\subfigure{\includegraphics[width=0.30\textwidth]{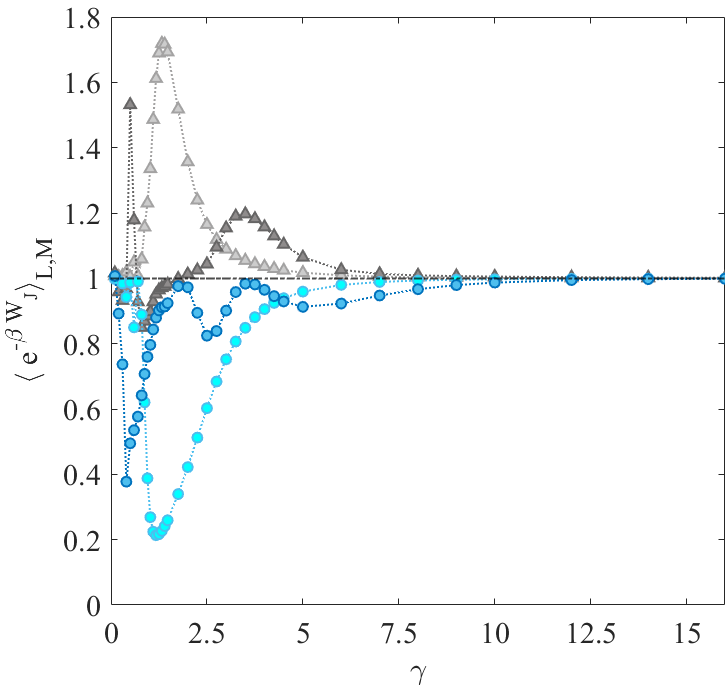}}	
    	\caption{Behaviour of $\ensavg_{L,M}$ as a function of the forcing frequency $\gamma$  for the coupled oscillators model, simulated with $\tau$ such that $B=\sin(2\pi)$. Left, center and right panel report the system behavior at varying values of the parameters $\beta$, $k_S$ and $k_I$, respectively. Markers represent results from numerical simulations, while dotted lines connecting them are a guide for the eye. Left panel: dark blue, light blue, light grey and dark grey lines correspond to $\beta=0.1,\;1,\;10,\;100$, respectively; other parameters are set to $m_S=m_E=1$, $k_S=k_E=1$, $k_I=1$ and $L=M=1$. Center panel: dark blue, light blue, light grey and dark grey lines correspond to $k_S=0.1,\;1,\;10,\;100$, respectively; other parameters are set to $m_S=m_E=1$, $k_E=1$, $k_I=1$, $\beta=1$ and $L=M=1$. Right panel, dark blue and light blue lines refer to $k_I=10$ and $k_I=0.1$, respectively, with $L=M=1$; dark grey and light grey plots refer to $k_I=10$ and $k_I=0.1$, respectively, but for $L=M=5$; remaining parameters are set to $m_E=10$, $m_S=1$, $k_S=k_E=1$, $\beta=1$.}
    \label{fig:couposcill_avexpWj_varK}
\end{figure}
\FloatBarrier
The pair of oscillators S and E from Sec. \ref{pairOS}, with periodic forcing  on S, shows a similar behavior, at least when the coupled particles have the same mass, $m_S=m_E=1$, and the interaction stiffness is sufficiently low (like $k_I=1$). This is illustrated by the first two panels of Fig. \ref{fig:couposcill_avexpWj_varK}, where the only parameters varied are $\beta$ and $k_S$. It is interesting to note how the temperature of the system needs to decrease (thus $\beta$ to increase) to make $\ensavg_{0;L,M}$ vanish. This is especially evident in central panel of Fig.\ \ref{fig:couposcill_avexpWj_varK} where none of the tested values of $k_S$ yields $0$ for $\beta=1$. On the contrary, $\beta=100$  obtains $\ensavg_{L,M}=0$ for $k_S=1$ and different values of $\gamma$. The reason is that higher $\beta$ implies a narrower distribution, hence analogous to a case with smaller $L$ and $M$.

Note that this is relevant also for infinite baths. A small temperature causes kinds of finite size effects, due to the smallness of the distribution variance. That, assuming the infinite space can at least in principle be explored, may be eliminated only at the cost of collecting enormous statistics, which is often impossible. Therefore, the finite ensemble result remains the only physically relevant.

The right panel of 
Fig.\ \ref{fig:couposcill_avexpWj_varK} shows the quantity \eqref{LM0} computed on a variation of the coupled oscillators model, in which the ``environment" mass $m_E$ is ten times bigger than the ``system" mass $m_S$.
%, to simulate a situation where the system exchanges energy with a bigger entity.
To include possible effects due to the efficiency of the energy exchange between system and environment, the stiffness of the interaction potential is varied: first a value of $k_I=0.1$ is implemented, then $k_I=10$ is employed to account for a rapid exchange of energy between the parts, that result almost rigidly connected. The figures show that the two configurations generate similar results when the initial ensemble is restricted to $L=M=1$, while noticeably different behavior is observed for a larger system, where $L=M=5$. The rigidly coupled system produces oscillations of $\ensavg_{0;L,M}$ around the resonance frequency, with the loosely connected case exhibiting even more evident down-peaks in the values of $\ensavg_{L,M}$ about the resonance frequency, as in the case of the periodically forced single oscillator. Moreover, %unlike all other cases investigated by us, 
both $L=5=M$ lead to peaks that exceed 1. Both weakly and strongly coupled oscillators indicate that the presence of a massive environment drastically magnifies the finite size effects, noticeably deviating from the equality $\ensavg_{0;L,M}=1$.

\section{Irreversible Expansion of an Ideal Gas}
    \label{s:idealgas_box_expansion}
    \par\noindent
     
\begin{figure}
         \centering
         \subfigure[]{\includegraphics[width=0.4\textwidth]{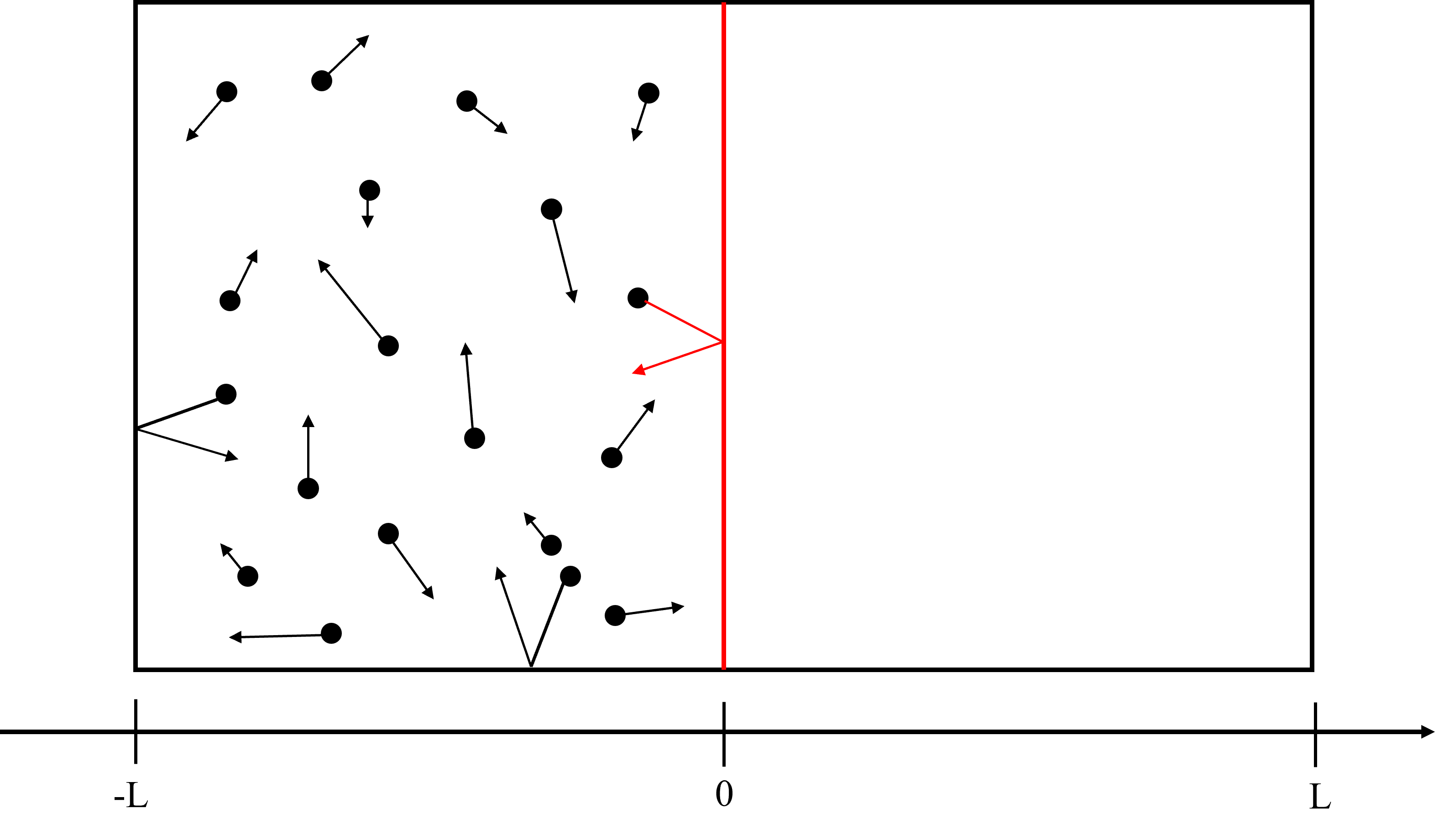}}\hspace{1cm}
         \subfigure[]{\includegraphics[width=0.4\textwidth]{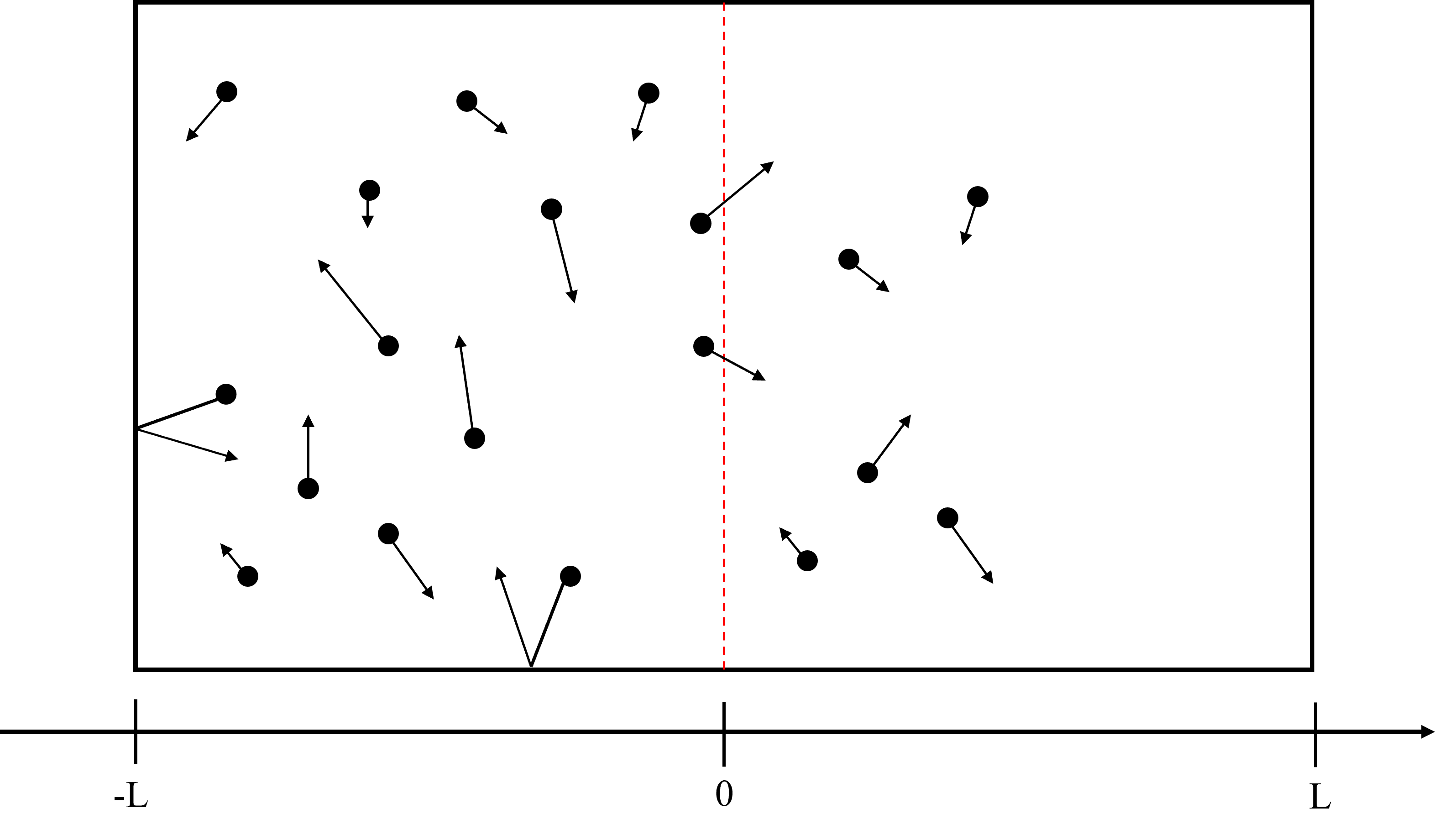}}\\
         \subfigure[]{\includegraphics[width=0.4\textwidth]{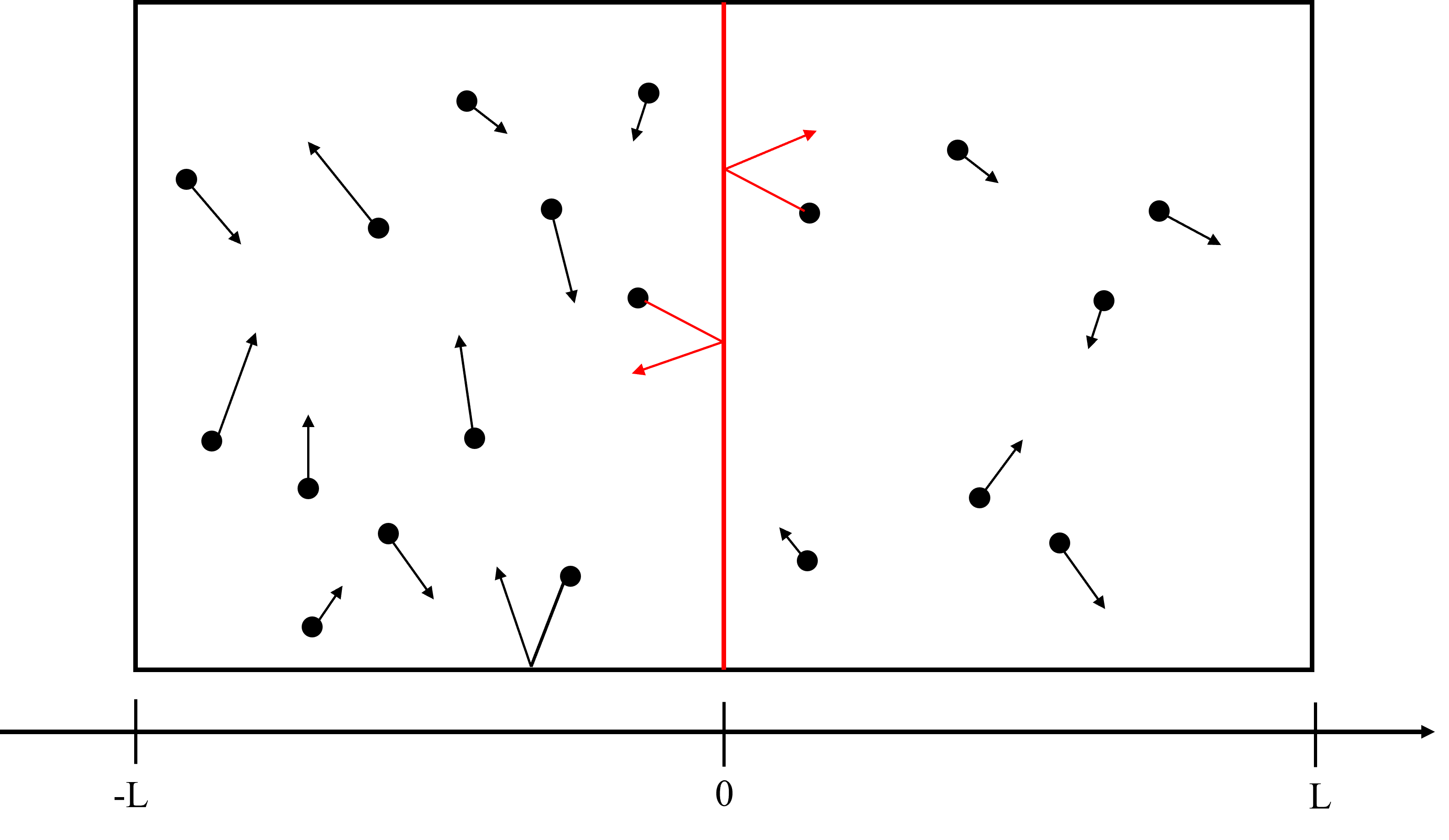}}
         \caption{Schematic representation of the irreversible expansion of an ideal gas. A 2D box contains non-interacting particles in equilibrium with a heat bath at temperature $T$. Panel (a): state of the system before the central wall is removed; all the particles are confined in the left half of the box and undergo specular reflections with the container walls. Panel (b): dynamics of the particles once the central wall is removed. Panel (c): after the wall is reintroduced, a fraction of particles is trapped in the right half of the box.}
         \label{fig:Gas_in_box}
\end{figure}

\begin{figure}[!ht]
	\centering
    \subfigure{\includegraphics[width=0.4\textwidth]{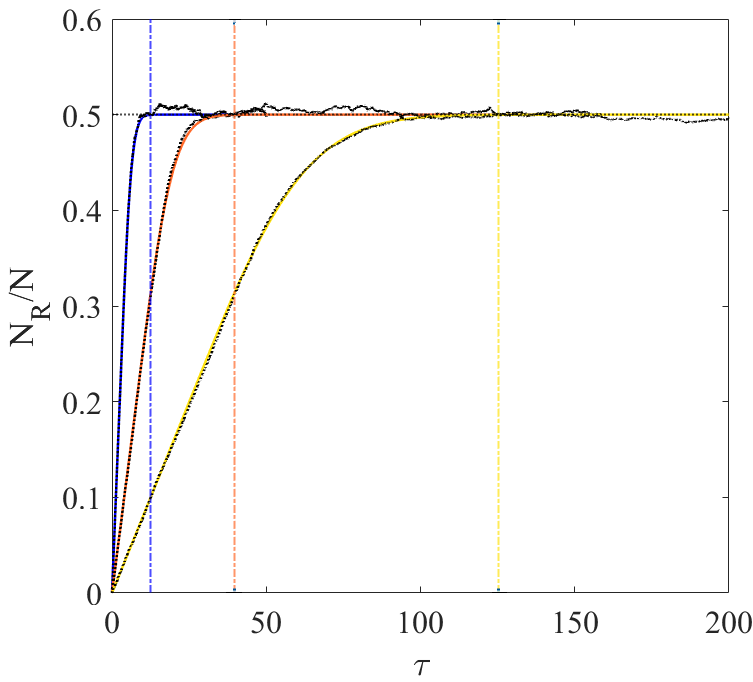}} \hspace{1cm}
    \subfigure{\includegraphics[width=0.4\textwidth]{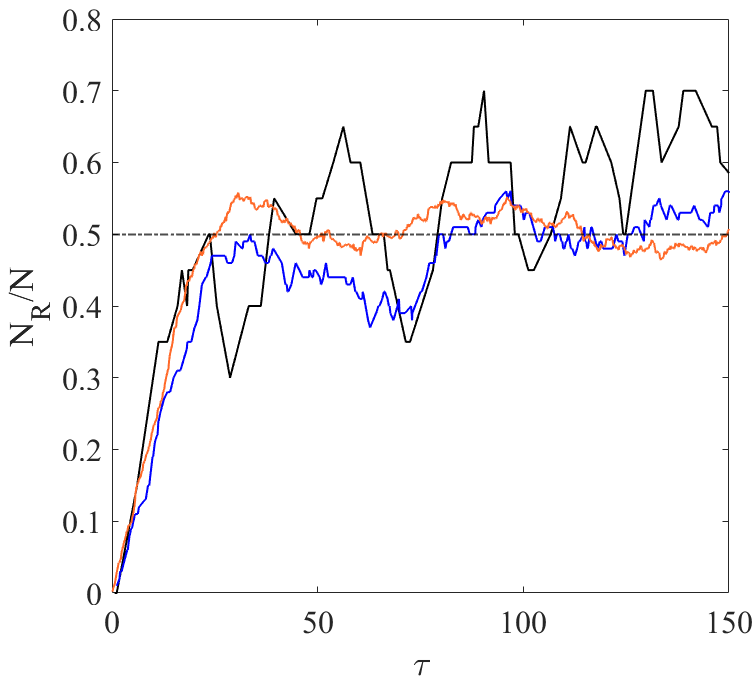}}
%\subfigure{\includegraphics[width=0.425\textwidth]{Chi_wh_mcTime_varBETA.png}}
\caption{\textit{Left panel}: behavior of $N_R/N$ as a function of the protocol duration $\tau$, for different values of $\beta$, with 
$N=10000$, $L=5$, $m=1$. Blue, orange and yellow solid lines refer to Eq.\eqref{eq:N_R_2} for $\beta=1,\;10,\;100$, respectively, and are obtained by truncating the formula \eqref{eq:N_R_2} to $n=1000$. The black dotted lines denote the results of the numerical simulations. Dash-dotted vertical lines indicate the characteristic times obtained from Eq. \eqref{eq:average_collision_time}.
\textit{Right panel}: behavior of $N_R/N$ vs. $\tau$, for different values of $N$, obtained from numerical simulations. Black, blue and orange solid lines refer to $N=20,\;100,\;500$ respectively. The other parameters are fixed to $L=5$, $\beta=10$, $m=1$.}
\label{fig:theory_check_1DGas}
\end{figure}

Consider a set of $N$ identical point particles in a 2D rectangular box of length $2L$.
The particles move in straight lines, and collide elastically with the hard boundaries of the box. The box is subdivided in two equal parts by a wall perpendicular to two sides, that can can be removed and placed back, according to prescribed protocols. 
We perform a cycle, starting from the gas confined by the wall in the left half of the box, and in equilibrium at a given temperature $T$. At time $t=0$, the system is isolated from the bath and the wall is removed for a certain amount of time $\zt$. Finally, the wall is placed back.
%     The dynamics of the system plus moving wall is straightforward: a number of particles, initially confined by the moving wall to the left half of the box, is free to evolve up to a certain time $t_1$, when the moving wall is removed; at this point, particles are able to escape the left half of the box and part of them will end up occupying the right half of the box. After a well defined time $\tau$, the wall is placed back in the box. 
A schematic representation of the system and its dynamics is given in Fig.\ \ref{fig:Gas_in_box}. The main observable here is the fraction of particles trapped in the right half of the box at the end of the cycle. Because this fraction depends on the details of the process through which the intermediate wall is removed and reinserted, the final equilibrium state may differ from the initial one, and may depend on the protocol.  
\FloatBarrier
For sake of simplicity, and  without any loss of significance, because the particles do not interact, we may replace the 2D container with a straight line segment of length $2L$. We also assume the particles to start with random  initial positions, uniformly distributed in the interval $(-L,0)$, and with initial velocities normally distributed, with mean $\mu=0$ and standard deviation $\sigma=\sqrt{{k_B T}/{m}}$, where $k_B$ is the Boltzmann constant, $T$ is the temperature of the bath and $m$ is the mass of the particles. 
%A negative (positive) velocity corresponds to an initially leftward (rightward) directed particle.
The fraction of particles escaping from the left half of the box to the right half, in the time interval $[0,\tau]$, is obtained integrating over all initial positions $x_0$ and initial velocities $v_0$ the probability for a particle to move from $(-L,0)$ to $(0,L)$. In the limit of many particles, the number of those leaving the left half and reaching the right half is this probability multiplied by their total number $N$. We denote by $N_L$ the number of particles in the left half of the box, and by $N_R$ the number of those in the right half of the box, so that $N=N_L+N_R$.
Now, note that the initial velocities pointing rightward (i.e. $v_0>0$) that make a particle starting at $x_0\in(-L,0)$ end in $x_{\tau}\in(0,L)$ after a time $\tau$, fulfill the inequalities:
\be
{1\over\tau}(4nL-x_0)<v_0<{1\over\tau}[(4n+2)L-x_0] \, , \quad n=0,1,2,\dots 
\label{vr}
\ee 
because travelling a distance of $(4nL-x_0)$ brings the particle in the interval $(0,L)$, after a number $n$ of bounces against the left wall of the container. Going beyond $[(4n+2)L-x_0]$ brings the same particle back to the left half of the  box. The same reasoning, applied to particles initially pointing leftward (hence $v_0<0$), shows that a particle is found in the right half of the box at time $\tau$ if its velocity $v_0$ is such that:
\be
%{1\over\tau}[(4n+2)L+x_0)<v_0<{1\over\tau}[(4n+4)L+x_0] \, , \quad n=0,1,2,\dots \quad \;.
-{1\over\tau}[(4n+4)L+x_0]<v_0<-{1\over\tau}[(4n+2)L+x_0) \, , \quad n=0,1,2,\dots 
\label{vl}
\ee
The number of particles residing in the right half of the box at time $\tau>0$, denoted as $N_R (\tau)$, is thus obtained by integrating over all initial positions $x_0$ in the interval $(-L,0)$ and over all initial velocities $v_0$ contained in the intervals given in \eqref{vr} and \eqref{vl}. By doing this, one implicitly assumes the system is made of infinitely many particles, therefore the result applies only for large $N$. For sufficiently large $N$, the following:
%Considering all the velocities, and 
%multiplying the corresponding probabilities by $N/2$, we finally obtain $N_R$, the number of particles in the right half of the box at time $\zt$, for assigned $L$ and $\sigma$:
    \be
%    N_R(\tau)={N\over L} \int_{-L}^0 dx \left\{ \sum_{n=0}^\infty \left[ {1\over 2} \int_{{1\over\tau}(4nL-x)}^{{1\over\tau}[(4n+2)L-x]} \mbox{d}v_+ {2 \over {\sigma\sqrt{2\pi}}}e^{-v_+^2/2\sigma^2} + {1\over 2} \int_{{1\over\tau}[(4n+2)L+x]}^{{1\over\tau}[(4n+4)L+x]} \mbox{d}v_- {2 \over {\sigma\sqrt{2\pi}}}e^{-v_-^2/2\sigma^2} \right]\right\}
    \frac{N_R(\tau)}{N}={1 \over {L\sigma\sqrt{2\pi}}} \int_{-L}^0 dx_0 \left\{ \sum_{n=0}^\infty \left[  \int_{{1\over\tau}(4nL-x_0)}^{{1\over\tau}[(4n+2)L-x_0]} e^{-v_0^2/2\sigma^2}\mbox{d}v_0  +  \int_{{1\over\tau}[(4n+2)L+x_0]}^{{1\over\tau}[(4n+4)L+x_0]} e^{-v_0^2/2\sigma^2}\mbox{d}v_0 \right]\right\}
    \label{eq:N_R_1}
    \ee
is thus an accurate prediction for observations.
Then, integrating first over the velocity space, exchanging the integral over positions with the infinite sum (which is made possible since each term of the infinite sum is a continuous, bounded function), and finally integrating over initial positions, Eq. \eqref{eq:N_R_1} yields
\bea
\begin{split}
    \frac{N_R(\tau)}{N} &=& {1\over2L}\sum_{n=0}^{\infty}\left\{ \sqrt{\frac{2}{\pi}}\sigma \tau %{S\over\sqrt{\pi}}
    \left[
    \exp \left( -{ 16 L^2 
    (n+1)^2 \over 2\sigma^2\tau^2} \right)
    + \exp \left( -{ 16 L^2 n^2 \over 2\sigma^2\tau^2 }   \right)
 -2 \exp \left( -{L^2 (4n+2)^2 \over 2\sigma^2\tau^2} \right) \right]\right.\\[5pt]
    && \left. \hskip 10pt - 2(4n+2)L \erf\left[{(4n+2)L\over \sqrt{2}\sigma\tau}\right]+4nL \erf\left[{4nL\over \sqrt{2}\sigma\tau}\right] 
      + (4n+4)L \erf\left[{(4n+4)L\over \sqrt{2}\sigma\tau} \right] \right\}
      \end{split}
    \label{eq:N_R_2}
    \eea
%The fraction of particles escaping to the right half of the container is obtained by simply dividing both sides of \ref{eq:N_R_2} by $N$. 
%where, to simplify notation, we introduced $S=\sqrt{2}\sigma\tau$. 
A comparison between this analytical result and numerical simulations of the system is shown in Fig.\ \ref{fig:theory_check_1DGas}. To estimate the relaxation times, in the absence of the intermediate wall, one cannot count on the mean time between two consecutive collisions with the boundaries of the box, given by $2 L \langle 1 / v \rangle$, because such a mean does not exist in a 1-dimensional space.
However, one may take the  distance $2L$ divided by the mean speed $\langle |v| \rangle$
    \be
    \widehat{t}={2L\over\langle |v| \rangle}=\sqrt{2\pi}{L\over\sigma}
    \label{eq:average_collision_time}
    \ee
as a characteristic time for the dynamics. This quantity estimates the time scale of the relaxation to a uniform distribution of particles, when their number is sufficiently high that recurrence times can be considered infinite to all effects.
As indicated by the vertical lines in the left panel of
Fig.\ \ref{fig:theory_check_1DGas}, systems with larger $\beta$, {\em i.e.}\ smaller temperature, take longer to reach the uniform distribution of particles in the box, as expected.
%In general, the mean collision times as defined above seem to give an indication of the time scale separating the spread of particles through the system (i.e., $\tau\leq\langle t \rangle$) from the steady condition where each half of the 1D box contains about half of the particles.
\FloatBarrier 
%
%    \begin{figure}[!ht]
%    	\centering
%    	\includegraphics[width=0.35\textwidth]{mean_coll_time_2definitions.png}
%    	\caption{\textbf{PLACEHOLDER}. Same as right panel in Fig.\ref{fig:theory_check_1DGas}, with black vertical line representing the characteristic  time $\widehat t$ \eqref{eq:average_collision_time}.}
%    	\label{fig:mean_collision_times}
%    \end{figure}
%    \FloatBarrier
The point here is that an irreversible process is generated by the motion of the moving wall, that returns to its initial position, at the end of a cycle. The result is that variations of the process time $\tau$ lead to different results for $N_R$, hence of the free energy of the final equilibrium states. 
For sufficiently large $\tau$, the process uniquely leads to $N_L=N_R$ (although fluctuations occur at any finite $N$, cf. the right panel of Fig. \ref{fig:theory_check_1DGas}), but that is different from the initial state $N_L=N$, $N_R=0$. Therefore, even accepting an ideal infinite thermostat, the variation of free energy between initial and final state does not vanish and depends on the process time $\tau$, at odds with the JE, {\em i.e.}\ with the canonical ensemble from which the JE is derived.
In this case, the canonical distribution of momenta is given by a Gaussian. Were the range of momenta finite, further corrections to the canonical prediction would arise.

For the dynamics to be  Hamiltonian, as the derivation of the JE requires,
the wall could be modelled by a repulsive potential $\Phi$, placed in the center of the box, that diverges at time 0 and $\tau$, and that vanishes at time $\tau/2$. For instance, the following would do:
\be
\Phi(x;\zl) = 
 \left\{ \begin{array}{lr}
0 & \mbox{if } ~ x \notin \left[ 1/2 - \epsilon , {1}/{2} + \epsilon \right] \\
\cfrac{1}{\zl} - \cfrac{4}{\zt^2} &  \mbox{if } ~ x \in \left[ 1/2 - \epsilon , {1}/{2} + \epsilon  \right] 
\end{array}
\right. \quad
\mbox{with } ~~
\zl(t) = t ( \zt - t )
\ee
with $2\epsilon>0$ representing the thickness of the wall.
In this case, the infinitesimal contribution to $W_J$, given by the interaction of particle $i$ in position $x_i$ with the wall, for a time ${\rm d} t$, is given by
\be
{\rm d} w_i = \left\{ \begin{array}{cr}
0 &  \mbox{if } ~ x_i \notin \left[ 1/2 - \epsilon , {1}/{2} + \epsilon \right] \\[4pt]
\cfrac{2 t - \tau}{t^2 ( \tau - t)^2} ~ {\rm d} t  &  \mbox{if } ~ x_i \in \left[ 1/2 - \epsilon , {1}/{2} + \epsilon  \right] 
\end{array}
\right. 
\ee
which has to be integrated over the time intervals within $[0,\tau]$ such that $x_i \in \left[ 1/2 - \epsilon , {1}/{2} + \epsilon  \right]$, and summed over all particles. 
Now, the standard canonical formalism is not applicable to this simple example, because the phase space corresponding to the initial equilibrium state is altered when the wall potential is lowered to finite height. It switches from representing an equilibrium state in half the volume of the box, to a different equilibrium state, that occupies the whole box. The instant in which the particles are allowed to move in whole the box, but are still confined in its left half, the initial canonical distribution does not describe anymore their state, and cannot be used to compute the free energy difference between the equilibrium states before and after the wall is removed. Nevertheless, this difficulty can be overcome, without affecting the result, considering, as generally and efficiently done (see {\em e.g.}\ Section 1.3 of Ref.\ \cite{kubo2012statistical}), 
that physically relevant space and time scales are finite, although they can be taken as large as one wants. Then, one realizes that a finite but high barrier confines a finite number of particles initially in the left half of the box, with only a negligible fraction $\epsilon$ of them moving to the other half, for a given time. A finite, but higher barrier confines the particles with same tolerance $\epsilon$ for a larger time, or for the same time and a smaller $\epsilon$. Given the (arbitrarily large) time and the (arbitrarily small) tolerance considered physically satisfactory, there is a barrier height that produces better confinement, allowing the initial state to be considered an equilibrium state. 
Then, the protocol dependence of the free energy difference 
described above remains.

\section{Concluding remarks}
	\label{s:discussion}
	\par\noindent
In this work we have discussed simple examples concerning finite size effects and a broken time reversal symmetry on the calculations of values of observables within the statistical mechanics formalism. 
It is indeed ever more important to properly describe systems that do not lie within the traditional bounds of statistical mechanics, developed for macroscopic systems at equilibrium, or slowly evolving near equilibrium states. Indeed, contemporary research widely focuses on small and far from equilibrium systems. In the case of equilibrium macroscopic systems, the use of the standard ensembles is fully justified, because the observables of interest are determined by the bulk, not the tails of the probability distributions. This approach is validated both by theory and experiments. 
On the contrary, fluctuations of properties of interest in the case of small or strongly 
nonequilibrium systems, compare to average signals, and require a proper characterization of the tails of the distributions, which may also be affected by a spontaneous time reversal symmetry breaking, not evident in the equations of motion. Indeed, the interaction with heat reservoirs is often limited to processes that last very short times, making effective only small parts of such environments. 
%To illustrate these points, we have first discussed the case of full versus truncated Gaussian distributions, and that of a random walk on a line, which is exactly solvable. We have shown that:
%\begin{itemize}
%    \item the finite size can be neglected only for observables that do not explode too fast in the tails of the distributions. From this point of view, exponential variables are problematic even if the tails of the probability distributions decay as fast as Gaussians.
%\end{itemize}
To illustrate these points we have investigated simple driven harmonic oscillators systems, averaging the popular quantity $\exp(-\beta W_J)$ with canonical and truncated canonical averages. %According to standard calculations based on the canonical ensemble, this quantity affords an estimate of the free energy difference $\Delta F$, between two equilibrium states at same temperature $T$, but characterized by two different values of a given parameter $\zl$, that occurs in the Hamiltonian of interest.
We have shown that:
\begin{itemize}
\item  
A single oscillator pulled by a constant speed harmonic trap yields the infnite bath result if the process is not too fast. It sensibly and rapidly departs from that when the speed of the driving agent grows.
The effect is more evident (as expected) for small than for large integration bounds, $L$ and $M$, for smaller harmonic constant, and for smaller bath temperatures. In the infinite $L,M$ limits, the standard canonical result is recovered, but larger and larger $L$ and $M$ are required, the smaller are $k_p$ or $\beta$.
\item For a single periodically driven oscillator, the infinite bath result over a multiple of the driving period equals 1. Strong deviations from this value, that even reach $0$, are found for a finite bath in finite intervals about the resonance frequency. While the theoretical result is again obtained in the infinite $L,M$ limit, this is harder if the driving acts for longer times ({\em i.e.}, for a larger number of periods).
\item In the case in which the oscillator S is coupled to an oscillator E, the infinite bath value 1 is obtained, apart from oscillations, for sufficiently large driving frequency. Noticeable deviations from that  results are still present about specific values of the forcing frequency. In this case, we have no analytical expression for the finite bath result. Therefore this conclusion is based on numerical data, for a finite ensemble of initial conditions, proven robust against variations of ensemble size.
\end{itemize}
The last example we have considered implies the breaking of the classical time reversal symmetry and consists of an ideal gas initially in equilibrium with an infinite thermal bath at temperature $T$, which is confined in the left half of its container by a moving barrier. Initially, the barrier confines all particles to the left half of the container, and that allows an equilibrium
state, represented by a uniform distribution for the positions of particles in $[-L,0]$ and a Gaussian distribution for their velocities. As soon as the barrier allows particles to reach the right half of the container the phase space changes to one in which positions cover the $[-L,L]$ interval, and the initial ensemble immediately fails to represent an equilibrium state. The observables take instead some time to change. This prevents the application of the standard statistical mechanical formalism, because the phase space of the equilibrium initial state is not the one of the nonequilibrium evolution and, for instance, the Liouville equation fails. 
A modification in time of the volume of a given system  to the very least requires a suitable time dependent scaling of the phase space coordinates \cite{ciccotti2021jarzynski}, for the formalism to apply, but that is not  possible in our case, because the volume changes instantaneously. Nevertheless, 
the experiment can be performed, and a proper formalism for it has been identified, in terms of finite potential barriers and time scales. 
%Then, given the value of $N_R$ at time $\tau$, there is a corresponding equilibrium at temperature $T$, whose free energy depends on $N_R$.
%Such an equilibrium can be described by a canonical ensemble that typically differs from the initial canonical ensemble, giving a protocol dependent $\Delta F$, that can be correctly evaluated in experiments or simulations, but not by a direct canonical calculation. 

Discrepancies between the canonical formalism and experimental situations are known to arise when  irreversibility emerges: they are intrinsic and not merely due to  
insufficient statistics. All boils down to the conclusion that finite size and irreversibility effects similarly lead to protocol dependent averages of exponential quantities such as $\exp(-\beta W_J)$. The  standard statistical mechanics formalism should be adapted to treat these cases.
This fits nicely with 
the standard statistical mechanical justification of ensembles. For instance, in Ref.\ \cite{Fermi}, Fermi states:\vskip 5pt
\noindent
{\em Studying the thermodynamical state of a homogeneous fluid of given volume at given temperature [...] we observe that there is an infinite number of states of molecular motion that correspond to it. With increasing time, the system exists successively in all the dynamical states that correspond to the given thermodynamical state. From this point of view we may say that a thermodynamical state is the ensemble of all the dynamical states through which, as a result of the molecular motion, the system is rapidly passing.}
\vskip 5pt
\noindent
and Callen adds 
\vskip 5pt
\noindent
{\em If the transition mechanism among the atomic states is sufficiently effective, the system passes rapidly through all representative atomic states in the course of a macroscopic observation [...].
%; such a system is in equilibrium
However, under certain unique conditions, the mechanism of atomic transition may be ineffective and the sytem may be trapped in a small subset of atypical atomic states. Or, even if the system is not completely trapped the rate of transition may be so slow that a macroscopic measurement does not yield a proper average over all possible atomic states. %In thsese cases the system is not in equilibrium.
}
\vskip 5pt
\noindent
In reality, less than required by Fermi and Callen is needed, for ensembles to work, because observables of interest are generally a few and well behaved \cite{Khinchin,marconi2008fluctuation,R-intro}. But when the standard conditions are severely violated, and the observables call for an accurate representation of large fluctuations, canonical results must be taken with a grain of salt.

\vskip 12pt
\noindent
{\bf Acknowledgements.}
LR acknowledges financial support by the Ministero dell'Universit\'{a} e della ricerca (Italy),
grant Dipartimenti di Eccellenza 2018 -- 2022 (E11G18000350001). This research was performed under
the auspices of Italian National Group of Mathematical Physics (GNFM) of INdAM.
	
	\appendix
	\section{Some explicit calculations for Sec. \ref{ss:linear_forcing}}
 \label{app:app1}

Retaining the notation of Sec. \ref{s:TheJE}, we denote by $P_0(\Gamma)$ the canonical distribution referring to a specific configuration $\Gamma$ (system $+$ environment), 
\be
    P_0(\Gamma)={1 \over Z_0} e^{-\beta {\cal H}(\Gamma; A)} \, , \quad \beta = \cfrac{1}{k_{_B} T} \; ,
    \label{CanEns}
    \ee
    with $k_B$ the Boltzmann's constant. One thus readily finds:
	\bea
	&&\hskip -55pt\Big\langle e^{-\beta W_{J,\ell}} \Big\rangle_0 = 
	{1 \over Z_0} \int_{-\infty}^\infty \int_{-\infty}^\infty \mbox{d}x_0 \mbox{d}p_0 e^{-\beta W_{J}(\ell;x_0,v_0)} e^{-\beta {\cal H}(x_0,v_0;0)}  =\\
	&&= \cfrac{1}{Z_0}
 \exp \left\{-\beta \frac{k_D k_p B^2}{2 k} - \beta {k_D^2 \ell^2 m\over k^2} \left( 1 - \cos \omega B/\ell \right) \right\}  \times \nonumber \\
	&&\hskip 10pt\int_{-\infty}^\infty \int_{-\infty}^\infty \mbox{d} x_0 \mbox{d} p_0 \exp \beta \left[ 
	{x_0 k_D \ell \over \omega} \sin \omega  {B \over \ell} -
	{p_0 k_D \ell \over k} \left( \cos \omega  {B \over \ell} - 1 \right)
	- {p_0^2 \over 2m}  - {k x_0^2 \over 2} \right]
	\label{aveWJl}
	\eea
	where $Z_0$ is the partition function of the initial canonical distribution and it is given by
	\be
	Z_0 = 
	\int_{-\infty}^\infty \mbox{d} x \, e^{-\beta {k x^2 / 2}} 
	\int_{-\infty}^\infty \mbox{d} p e^{-\beta {p^2 / 2 m}} = {2 \pi \over \beta\zw}.
	\label{InCan}
	\ee
	The double integral in \eqref{aveWJl} can be separated and computed in two parts:
	\be
	\int_{-\infty}^\infty \mbox{d} \, x \, e^{\beta x
		{k_D \ell \over \omega} \sin \omega  {B \over \ell} - {\beta k \over 2} x^2}
	= \sqrt{2 \pi \over \beta k} \exp \left( {\beta k_D^2 \ell^2 m\over 2 k^2} \sin^2 \omega  {B \over \ell} \right)
	\label{positave}
	\ee
	and
	\be
	\int_{-\infty}^\infty \mbox{d} \, p \, e^{-\beta \left[ p
		{k_D \ell \over k} \left( \cos \omega  {B \over \ell} -1 \right) + { p^2 \over 2m} \right]
	} = \sqrt{2 \pi m\over \beta} \exp \left[ {\beta k_D^2 \ell^2 m \over 2 k^2} \left( 
	\cos^2 \omega  {B \over \ell} - 2 \cos \omega  {B \over \ell}  + 1 \right) \right] \; ,
	\label{velave}
	\ee
	from which it follows that:
	\bea
	&&\hskip -50pt
	\Big\langle e^{-\beta W_{J,\ell}} \Big\rangle_0 = {\beta \zw \over 2 \pi} {2 \pi \over \beta \zw}
	\exp\left\{-\beta {k_D k_p B^2 \over 2 k} - \beta {k_D^2 \ell^2\over k \zw} \left( 1 - \cos \zw B/\ell \right)
	\right\} \times \nonumber\\
	&& \hskip 15pt \exp \left[{\beta k_D^2 \ell^2\over 2 k \zw} \sin^2 \omega  {B \over \ell} +
	{\beta k_D^2 \ell^2 \over 2 k \zw} \left( 
	\cos^2 \omega  {B \over \ell} - 2 \cos \omega  {B \over \ell}  + 1 \right) 
	\right] \nonumber\\
	&&=
	\exp \left\{ -\beta 
	{k_D k_p B^2 \over 2 k} \right\}
	\label{aveWJlfin} \; .
	\eea

Let us now turn to consider canonical distributions truncated
at a given distance $L$ from the rest position of the oscillator, and at a maximum momentum $M$.
Referring to the model of a single oscillator subject to a linear protocol, treated in Sec. \ref{ss:linear_forcing}, we denote:
\be
P_0(x,p) = \cfrac{1}{Z_0(L,M)}
\left\{
\begin{array}{lcr}
 e^{-\beta {(k x^2 + p^2/m)/ 2}} & \mbox{if} & | x | \le L ~\mbox{ and }~ |p| \le M \\[7pt]
0  & \mbox{if} & | x | > L ~\mbox{ or }~ |p| > M  
\end{array}
\right.
\label{P0xp_bis}
\ee
where:
	\be
	Z_0(L,M) = 
	\int_{-L}^L \mbox{d} x \, e^{-\beta {k x^2 / 2}} 
	\int_{-M}^M \mbox{d} p e^{-\beta {p^2 / 2 m}} = {2 \pi \over \beta \omega} \,
	\mbox{erf}\left(\sqrt{\beta k \over 2} L \right) \mbox{erf}\left(\sqrt{\beta \over 2 m} M \right)
	\label{FinSiz}
	\ee
Consequently, the average of $\exp(-\zb W_J)$ for a given $\ell$ now reads:
\bea
	&&\hskip -60pt
 \Big\langle e^{-\beta W_{J,\ell}} \Big\rangle_{0;L,M} = \cfrac{1}{Z_{0;L,M}}
	\exp \left[-\beta {k_D k_p B^2 \over 2 k} - \beta {k_D^2 \ell^2 m\over k^2} \left( 1 - \cos \omega B/\ell \right) \right]
		\times \\[7pt] 
	&&\hskip -30pt\int_{-L}^L \int_{-M}^M \mbox{d} x_0 \mbox{d} p_0 \exp 
 \left\{\beta \left[ 
	{x_0 k_D \ell \over \zw} \sin \omega  {B \over \ell} -
	{p_0 k_D \ell \over k} \left( \cos \omega  {B \over \ell} - 1 \right)
	- {p_0^2 \over 2m}  - {k x_0^2 \over 2} \right] \right\}
	\label{aveFWJl}
	\eea
	where we can separately compute:
	\bea
	&&\hskip -50pt 
 \int_{-L}^L \mbox{d} \, x \, e^{\beta x
		{k_D \ell \over \zw} \sin \omega  {B \over \ell} - {\beta k \over 2} x^2}
	= \sqrt{\pi \over 2 \beta k} e^{\beta k_D^2 \ell^2 m \sin^2 \omega {B \over \ell} \over 2 k^2} \\
	&& \times \left[ \mbox{erf} \left( \sqrt{\beta k \over 2} L + \sqrt{\beta m \over 2} 
	{k_D \ell \sin \omega {B \over \ell} \over k} \right) 
	+ \mbox{erf} \left( \sqrt{\beta k \over 2} L - \sqrt{\beta m \over 2} 
	{k_D \ell \sin \omega {B \over \ell} \over k}
	\right)
	\right]
	\label{positaveF}
	\eea
	and
	\bea
	&&\hskip -65pt \int_{-M}^M \mbox{d} \, p \, e^{-\beta \left[ p
		{k_D \ell \over k} \left( \cos \omega  {B \over \ell} -1 \right) + { p^2 \over 2m} \right]} 
	= \sqrt{\pi m \over 2 \beta} e^{ {\beta k_D^2 \ell^2 m \over 2 k^2} \left(
		\cos \omega  {B \over \ell} - 1 \right)^2} \\
	&& \times \left[ \mbox{erf} \left( \sqrt{\beta \over 2 m} M - \sqrt{\beta m \over 2} {k_D \ell \over k} 
	\left( \cos \omega  {B \over \ell} - 1 \right)
	\right) \right. \\
	&& \hskip 85pt \left. + \, \mbox{erf} \left( \sqrt{\beta \over 2 m} M + \sqrt{\beta m \over 2} {k_D \ell \over k} 
	\left( \cos \omega  {B \over \ell} - 1 \right)
	\right) \right]  
	\label{velaveF}
	\eea
	Therefore, one obtains:
	\be
\Big\langle e^{-\beta W_{J,\ell}} \Big\rangle_{0;L,M} = %\cfrac{1}{Z_{0;L,M}}\cdot 
 I_{exp}\cdot I_x \cdot I_p
	\ee
	with
	\bea
	&&I_{exp} = \exp \left\{-\beta {k_D k_p B^2 \over 2 k } \right\} = 
 \Big\langle e^{-\beta W_{J,\ell}} \Big\rangle_0 \\[5pt]
	&&I_x = { \mbox{erf} \left( \sqrt{\beta k \over 2} L + \sqrt{\beta m \over 2} 
		{k_D  \over k}\ell \sin \omega {B \over \ell}  \right) 
		+ \mbox{erf} \left( \sqrt{\beta k \over 2} L - \sqrt{\beta m \over 2} 
		{k_D  \over k}\ell \sin \omega {B \over \ell}
		\right) \over 2 ~ \mbox{erf}\left(\sqrt{\beta k \over 2} L \right) } \\[5pt]
	&&I_p = { \mbox{erf} \left( \sqrt{\beta \over 2 m} M + \sqrt{\beta m \over 2} {k_D\over k} \ell  
		\left( \cos \omega  {B \over \ell} - 1 \right)
		\right) 
		+ \, \mbox{erf} \left( \sqrt{\beta \over 2 m} M - \sqrt{\beta m \over 2} {k_D \over k} \ell 
		\left( \cos \omega  {B \over \ell} - 1 \right)
		\right) \over 2 ~ \mbox{erf}\left(\sqrt{\beta \over 2 m} M \right) } 
	\eea

	\section{Some explicit calculations for Sec. \ref{ss:periodic_forcing}}
 \label{app:app2}

For the model of a single oscillator subject to a periodic forcing, discussed in Sec. \ref{ss:periodic_forcing}, one has:
	\bea
	&&\hskip -40pt \Big\langle e^{-\beta W_J} \Big\rangle_{0;L,M} = { 1 \over Z_0(L,M)} \times  \nonumber \\[5pt]
 &&\hskip -25pt 
 \exp\left\{ -{\beta k_D \over 4}\left( 1-{k_D/m \over \omega^2-\gamma^2}\right) \left( 1 - \cos 2\gamma \tau \right) - {\beta k_D^2\gamma^2/m \over (\omega^2-\gamma^2)^2}\left(1-{\gamma\over\omega}\sin\gamma\tau\sin\omega\tau  -\cos\gamma\tau\cos\omega\tau\right) \right\} \times \nonumber \\[5pt]
	%\vphantom{\frac12}\right]
	&& \int_{-L}^L \mbox{d} x \, 
 \exp\left\{-{\beta k \over 2}x^2 + {\beta k_D\gamma\omega \over \omega^2-\gamma^2} \left( \cos\gamma\tau\sin\omega\tau -  {\gamma\over\omega}\sin\gamma\tau\cos\omega\tau\right)x \right\} \times 
 \nonumber \\[5pt]
	&& \int_{-M}^M \mbox{d} p \, 
 \exp\left\{- { \beta \over 2 m}p^2 + {\beta k_D\gamma/m\over\omega^2-\gamma^2}\left(1-{\gamma\over\omega}\sin\gamma\tau\sin\omega\tau-\cos\gamma\tau\cos\omega\tau\right)p
  \right\}
	\label{av_exp_WJ_expanded}
	\eea
which thus leads to:
	\be
	\Big\langle e^{-\beta W_J} \Big\rangle_{0;L,M} = I_{exp} \cdot I_x \cdot I_p  
	\label{av_exp_WJ_explicit}
	\ee
where
	\bea
	&&\hskip -50pt I_{exp}=\exp \left\lbrace -{\beta k_D \over 4}\left( 1-{k_D/m \over \omega^2-\gamma^2}\right) \left( 1 - \cos 2\gamma \tau \right)  \right.\\[5pt]
	&& \hskip 20pt \left. -{\beta k_D^2\gamma^2 \over (\omega^2-\gamma^2)^2} \left[ {1\over m} \left(1-{\gamma\over\omega}\sin\gamma\tau\sin\omega\tau  -\cos\gamma\tau\cos\omega\tau\right)  \right. \right.\\[5pt]\\
	&& \hskip 20pt \left. - {1\over2m} \left(1-{\gamma\over\omega}\sin\gamma\tau\sin\omega\tau  -\cos\gamma\tau\cos\omega\tau\right)^2 \right. \\[5pt]
	&& \hskip 40pt \left.  - {\omega^2 \over 2k}\left(\cos\gamma\tau\sin\omega\tau - {\gamma\over\omega}\sin\gamma\tau\cos\omega\tau\right)^2\biggr]  \vphantom{\frac12}\right\rbrace
	\label{Iexp}
	\eea
	
	\bea
	&& I_x= {1\over2 ~ \mbox{erf}\left(\sqrt{\beta k \over 2 } L \right)} \left[ \mbox{erf}\left({\beta k L - {\beta k_D \gamma \omega\over\omega^2-\gamma^2}\left(\cos\gamma\tau\sin\omega\tau-{\gamma\over\omega}\sin\gamma\tau\cos\omega\tau\right)\over \sqrt{2\beta k} } \right) + \right.\\
	&& \hskip 105pt \left. \mbox{erf}\left({\beta k L + {\beta k_D \gamma \omega\over\omega^2-\gamma^2}\left(\cos\gamma\tau\sin\omega\tau-{\gamma\over\omega}\sin\gamma\tau\cos\omega\tau\right)\over \sqrt{2\beta k} } \right) \right]
	\label{Ix}
	\eea
	
	\bea
	&& I_p= {1\over2 ~ \mbox{erf}\left(\sqrt{\beta  \over 2 m} M \right)} \left[ \mbox{erf}\left({{\beta\over m} M - {\beta k_D \gamma/m\over\omega^2-\gamma^2}\left(1-{\gamma\over\omega}\sin\gamma\tau\sin\omega\tau-\cos\gamma\tau\cos\omega\tau\right)\over \sqrt{2\beta /m} } \right) \right.\\
	&& \hskip 105pt \left. + \mbox{erf}\left({{\beta\over m} M + {\beta k_D \gamma/m\over\omega^2-\gamma^2}\left(1-{\gamma\over\omega}\sin\gamma\tau\sin\omega\tau-\cos\gamma\tau\cos\omega\tau\right)\over \sqrt{2\beta /m} } \right) \right] \; .
	\label{Ip}
	\eea

%	\label{s:appendixA}
%	\par\noindent

	\bibliographystyle{ieeetr}
	\bibliography{biblio_arxiv}
	
\end{document}